\documentclass[preprint,showpacs,preprintnumbers,amsmath, nofootinbib]{revtex4}
\usepackage{amsmath,amssymb,graphicx}
\usepackage{epstopdf}
\usepackage {amssymb}
\newcommand{\nc}{\newcommand}
\nc{\ba}{\begin{eqnarray}}
\nc{\ea}{\end{eqnarray}}
\newcommand\be{\begin{equation}}
\newcommand\ee{\end{equation}}
\newcommand\bse{\begin{subequations}}
\newcommand\ese{\end{subequations}}

\def\Tr{  \mbox{Tr}   }

\makeatletter \@addtoreset{equation}{section}

\makeatletter\renewcommand\section{\@startsection {section}{1}{\z@}%
                                   {-3.5ex \@plus -1ex \@minus -.2ex}
                                   {2.3ex \@plus.2ex}%
                                   {\normalfont\large\bfseries}}
\renewcommand\subsection{\@startsection{subsection}{2}{\z@}%
                                     {-3.25ex\@plus -1ex \@minus -.2ex}%
                                     {1.5ex \@plus .2ex}%
                                     {\normalfont\bfseries}}

\input{epsf.sty}

\begin{document}

\begin{flushright}\vspace{-2cm}
{\small
{\tt arXiv:0911.4284 [hep-th]} \\
IPM/P-2009/049\\
UUITP-27/09\\
MCTP-09-54} %
\end{flushright} 

\title{ Matrix Inflation and the Landscape of its Potential}

\author{Amjad Ashoorioon$^{1,2}$}
\email{amjad.ashoorioon(AT)fysast.uu.se}
\author{Hassan Firouzjahi$^{3}$}
\email{firouz(AT)ipm.ir}
\author{Mohammmad Mahdi Sheikh-Jabbari$^{3}$}
\email{jabbari(AT)theory.ipm.ac.ir}
\affiliation{1)
Michigan Center for Theoretical Physics, University of
Michigan, Ann Arbor, Michigan 48109-1040, USA}
\affiliation{2) Institutionen f\"{o}r fysik och astronomi
Uppsala Universitet, Box 803, SE-751 08 Uppsala, Sweden  }
\affiliation{3) School of Physics, Institute for Research in Fundamental Sciences (IPM),
P. O. Box 19395-5531, Tehran, Iran}


\begin{abstract}

\vspace{1cm}
Recently we introduced an inflationary setup in which the inflaton fields are
matrix valued scalar fields with a generic
quartic potential, M-flation. In this work we study the landscape of
various inflationary models arising from M-flation. The  landscape
of the inflationary potential arises from the dynamics of concentric
multiple branes in appropriate flux compactifications of string
theory. After discussing the classical landscape of the theory we
study the possibility of  transition among various inflationary
models appearing at different points on the landscape, mapping the
quantum landscape of M-flation. As specific examples, we study
some two-field inflationary models arising from this theory in the
landscape.

\vspace{1cm}

\end{abstract}

\maketitle

\tableofcontents

\section{Introduction}

The inflationary paradigm, the idea that the early Universe has
undergone a nearly exponential expansion phase, has appeared as the
leading candidate for explaining the recent cosmological
observations data \cite{Komatsu:2008hk}. The simplest and still
successful model of inflation is a massive, free scalar field
minimally coupled to the Einstein gravity. Nonetheless, motivated by
various beyond  the Standard Model particle physics or supergravity and
string theory settings, many models of inflation have been
constructed by introducing more non-trivial potentials for the
scalar field and/or the addition of other scalar fields to the model.

The option of having a large number of scalar fields, instead of a
handful of them, has been particularly motivated by string theory
inspired models, where after fixing the moduli in compactifications
they appear as light scalar fields subject to different potentials
\cite{HenryTye:2006uv}. A large number of (decoupled) scalars has
the advantage that the contribution of each of them to the Hubble
expansion parameter during inflation adds up, leading to a
substantial number of e-folds, even if the potential for each field
is not flat enough to sustain a successful period of inflation
\cite{Liddle:1998jc}. On the observational sides, a generic multiple
field inflationary system predicts entropic and non-Gaussian
perturbations which both are under intense observational constraints
\cite{Komatsu:2008hk}.

In \cite{Ashoorioon:2009wa}, again motivated by string and D-branes
settings, we introduced a model of
inflation which has a large number of  fields. In this model the inflaton fields were taken  to be matrix valued objects and hence the
model was dubbed as Matrix Inflation, or M-flation in short. The
non-Abelian and non-commuting nature of matrices play a crucial
role in M-flation construction.

We showed in \cite{Ashoorioon:2009wa} that in a special corner of
the rich ``landscape" of the M-flation, one can reduce the theory to
a standard chaotic inflationary model with at most a quartic
potential. The matrix nature, however, now shows up in removing (or
easing) the unnaturally small values of the couplings for these
models. In this work we would like to analyze the landscape
potential of M-flation in more detail. The landscape arises from the dynamics of
coincident branes  subject to appropriate fluxes in string theory
compactification. At different points of this landscape we have
inflationary models which are classically disconnected and the
parameters at each point could be adjusted to achieve a period of
slow-roll inflation with enough number of e-folds and acceptable
spectral index. These different vacua, in principle, can tunnel to each
other via quantum effects. If some fine-tunings of the couplings are tolerated, similar to the old inflation idea which should have terminated through first order phase transition \cite{Guth:1980zm}, slow-roll inflation could end through bubble collision.

The paper is organized as follows. After reviewing the basic setup of M-flation in section \ref{setup},
we study the landscape of M-flation theories in section
\ref{Theory-landscape}. In section \ref{inflation-landscape} we
analyze the landscape of the potential and hence inflationary models
for a given Matrix inflation theory. Here we give a counting of
various possible models (a counting on the landscape). This is
basically a counting of number of $N\times N$ reducible
representations of SU(2), which is  much larger than $N^2$ for large $N$. We
also discuss the tunneling between various vacua. As  specific
examples, we study some two-field inflationary models arising in our
setup. In section \ref{two-block}, we study numerically the physical
predictions of these two-field models for CMB perturbations.  The
conclusions and discussions are provided in section
\ref{discussions}. In  Appendix A we present the cosmological
perturbation theory for two field
models in some details.  In  Appendix B, we consider the case where we have a positive energy metastable vacuum and find out the set of couplings where the first order phase transition from metastable vacuum to the true one is possible via  quantum tunneling. As stated above, for a small window of parameters, nucleation rate is large enough to allow for a a first order phase transition via Coleman-De Luccia phase transition.

\section{M-flation, the Setup}
\label{setup}

As in  \cite{Ashoorioon:2009wa} we consider M-flation the inflationary model in which the inflaton fields are considered to be matrices and
the inflationary potential is constructed from the matrices and their commutators. The action is
\ba \label{action}%
 S=\int d^{4} x \sqrt {-g} \left(\frac{ M_{P}^{2} }{2} R - \frac{1}{2}
\sum_{i} \Tr  \left( \partial_{\mu} \Phi_{i} \partial^{\mu} \Phi_{i}
\right) - V(\Phi_{i}, [ \Phi_{i}, \Phi_{j}] ) \right) \, , %
\ea %
where the reduced Planck mass is $M_{P}^{-2}= 8 \pi G$ with $G$
being the Newton constant and the  signature of the metric is
$(-,+,+,+)$.  Here the index $i$ counts the number of matrices and
we take it to be $i=1,2,3$. The kinetic energy of $\Phi_i$ has the
standard form and $\Phi_i$ are minimally coupled  to gravity.
To be specific, as in  \cite{Ashoorioon:2009wa}, we consider the following potential
\ba\label{The-Potential}%
V= \Tr  \left( - \frac{\lambda}{4}  [ \Phi_{i},
\Phi_{j}] [ \Phi_{i}, \Phi_{j}] +\frac{i \kappa}{3} \epsilon_{jkl}
[\Phi_{k}, \Phi_{l} ] \Phi_{j} +  \frac{m^{2}}{2}  \Phi_{i}^{2}
\right) \,,%
\ea%
which is quadratic in $\Phi_i$ or $[\Phi_i,\Phi_j]$; \eqref{The-Potential} is the most general potential with this property.
We choose  the three parameters $\lambda,\ \kappa$ and $m^2$ to be non-negative.

The above action enjoys a global $U(N)$  symmetry under which
$\Phi_i$ are in the adjoint. One may also consider the theory in
which this $U(N)$ is gauged. The latter is done by replacing the partial
derivatives of the scalars $\Phi_i$ by covariant derivatives,
$D_\mu\Phi_i=\partial_\mu \Phi_i+ig[A_\mu,\Phi_i]$, where $A_\mu$ is
the gauge field, and by adding the Yang-Mills term for the gauge
fields. For our purposes, during (slow-roll) inflation and as far as
the classical dynamics of the scalars is involved, one may
consistently set the gauge fields to zero, in which case the
inflationary dynamics of the gauged and un-gauged models become
identical.
These two models, however, can have a different spectrum of linear
perturbations about the inflationary background.

In  \cite{Ashoorioon:2009wa} we argued that this system, and in
particular when the $U(N)$ symmetry is gauged, is strongly motivated
from string theory where $\Phi_i$ represents the low energy dynamics
of coincident branes in some specific flux compactifications
\cite{Myers:1999ps}.
 In this
picture, $N$ coincident D3-branes extended along our Universe,
subject to  RR six-form field $C_{(6)}$, and through the Myers
effect \cite{Myers:1999ps}, can blow up into D5-branes, wrapping
around a two-dimensional sphere in the extra three dimensions which
are parameterized by matrix valued scalars $\Phi_i$. \footnote{We
note that, as discussed in \cite{Ashoorioon:2009wa}, the action
\eqref{action} with the potential \eqref{The-Potential} is the
action  in the lowest order in string scale $\alpha'$. Considering
$\alpha'$ corrections adds term higher order in $\Phi_i$. Therefore
the string theory picture for our model is valid if these $\alpha'$
corrections are small.} This two-sphere for finite $N$ is a fuzzy
two-sphere and in the large $N$ limit it becomes a commutative round
sphere. In \cite{Ashoorioon:2009wa} we studied the case where
$N\times N$ matrices exhibit a single fuzzy two-sphere. It is
notable that in the large $N$ limit the sector of M-flation studied
in \cite{Ashoorioon:2009wa},  similar to \cite{Thomas:2007sj},
\cite{Ward:2007gs} and \cite{Berndsen:2009ww}, is closely connected
to the models of D5-branes wrapping two cycles of an internal
(Calabi-Yau) space \cite{Becker:2007ui,giant-inflaton}. In these
models the D5-branes are moving in a Klebanov-Strassler throat
\cite{KS} where the DBI effects become important. In this work,
however, we would like to focus on the matrix effects. As we show
below due to the matrix nature of the fields we have the possibility
of  multi-field inflationary models which does not exist in the
analysis of \cite{Becker:2007ui,giant-inflaton}.

As we will show in this work, there is also the possibility that we
have a large number of concentric fuzzy spheres of various radii,
corresponding to a picture where we have more than one fuzzy sphere
or wrapped D5-brane. Moreover, there is a distinct inflationary
model associated with each of these multi-fuzzy spheres solutions:
For the case of multi fuzzy spheres solution we obtain a multiple field
inflationary model, where these scalar fields are classically
decoupled from each other. These fields, similarly to the
single field of \cite{Ashoorioon:2009wa}, have the geometric
interpretation of radii of the fuzzy spheres. To see this
we analyze the equations of motion of the theory. Starting with an
isotropic and homogenous  FRW background
\be%
ds^{2} = -dt^{2} + a(t)^{2} d   \vec{\bf x}^{\, 2} \, ,
\ee%
the equation of motions are%
 \bse\label{eom}\begin{align}%
&H^{2}= \frac{1}{3 M^2_{P}} \left( - \frac{1}{2} \Tr
\left(
\partial_{\mu} \Phi_{i} \partial^{\mu} \Phi_{i}  \right) +
V(\Phi_{i}, [ \Phi_{i}, \Phi_{j}] ) \right) \\ &\ddot \Phi_{l} + 3 H
\dot \Phi_{l} + \lambda   \left[ \Phi_{j}, \, [\Phi_{l}, \Phi_{j}]
\,  \right] + i \, \kappa \, \epsilon_{l j k } [ \Phi_{j}, \Phi_{k}]
+ m^{2} \Phi_{l} =0 \, ,\\
&\dot H=-\frac{1}{2M^2_{P}}\ \sum_{i} \Tr
\partial_{\mu} \Phi_{i} \partial^{\mu} \Phi_{i}\ ,%
\end{align}\ese%
 where $H= \dot a/a$ is the Hubble expansion rate.

It is straightforward to check that the above equations of motion
can be solved with $\Phi_i$ in the matrix form%
\be\label{general-solution}%
\Phi_i=\sum_\alpha \hat\phi_\alpha J_i^\alpha \quad, \quad     i=1, 2,3       %
\ee%
where $J_i^\alpha$ are $N_\alpha\times N_\alpha$ matrices  satisfying the $SU(2)$
algebra%
\be%
[J_i^\alpha, J_j^\beta]=i\epsilon_{ijk} \delta^{\alpha\beta}\
J_k^\alpha\ ,\qquad Tr(J_i^\alpha J_j^\beta)=\frac14
N_\alpha(N^2_\alpha-1)\delta_{ij}\delta^{\alpha\beta}\ .%
\ee%
Here $N_\alpha$ are arbitrary non-negative integers subject to the
condition $\sum_\alpha N_\alpha=N$. That is, $\sum_\alpha
J_i^\alpha$ form reducible $N\times N$ representation of $SU(2)$. In
the above, the value of $\alpha$  specifies the number of
irreducible $SU(2)$ blocks in the generic $N\times N$ matrix.
Therefore, the range of $\alpha$ can vary from one, corresponding to
a single irreducible representation which was discussed in some
details in \cite{Ashoorioon:2009wa}, to $N$, where basically we have
the solution $\Phi_i=0$.

In fact, one can show that \eqref{general-solution} is the most
general solution to (\ref{eom}b). Moreover, sum of two
solutions of the form \eqref{general-solution} is not a solution to
(\ref{eom}b). With the above decomposition one finds  that the
equations for $\hat \phi_\alpha$ decouple and one may analyze them
separately. It is then convenient to rewrite the action in terms of
$\hat\phi_\alpha$, rather than the matrices
$\Phi_i$. Doing so, and after rescaling%
\be\label{phi-scaling}%
\hat\phi_\alpha=\left[\frac14
N_\alpha(N^2_\alpha-1)\right]^{-1/2}
\phi_\alpha\ ,%
\ee%
we obtain a multi-field canonically normalized action for the
scalars $\phi_\alpha$ with the potential%
\be\label{potential-generic}%
V(\phi_\alpha)=\sum_\alpha\ \frac{\lambda_{\alpha}}{4}
\phi_\alpha^{4}-
\frac{2\kappa_{\alpha}}{3} \phi_\alpha^{3}+\frac{m^{2}}{2} \phi_\alpha^{2} %
\ee%
where%
\be\label{lambda-kappa-scaling}%
 \lambda_{\alpha}= \frac{8 \lambda}{ N_\alpha (N_\alpha^{2}-1)} \
, \qquad \kappa_{\alpha} = \frac{2 \,
\kappa}{\sqrt{N_\alpha(N_\alpha^{2}-1)}}. %
\ee%
As we see the fields $\phi_\alpha$ are decoupled from each other,
which is in fact the result (or advantage) of using the $SU(2)$
basis we have introduced. Note that although in the potential
\eqref{potential-generic} there is no interaction between
$\phi_\alpha$, they all couple to gravity and contribute to $H$.
Therefore, their dynamics are coupled through gravity.

{}From now on we call each reducible solution which minimizes the
potential a ``vacuum'' solution. In \cite{Ashoorioon:2009wa} we
analyzed the  irreducible vacuum, where we studied the $N_\alpha=N$
case in which $\alpha=1$ and as such this vacuum may also be called
the ``single block vacuum'' or  the ``single giant vacuum'' (in a
reference to the blown-up D3 branes picture). Similarly, when
$\alpha$ ranges from one to $n$ we have an $n$-block vacuum or
$n$-giant vacuum.

\section{The  Landscape of M-flation theories}\label{Theory-landscape} %

There are four parameters in the original M-flation action,
$\lambda,\ \kappa,\ m^2$ and the size of matrices $N$. As discussed
above, the classical solutions of the theory are described by the set of
parameters $(\lambda_\alpha,\kappa_\alpha, m^2)$ and the set of
$N_\alpha$ which are subject to $\sum_{\alpha=1}^n N_\alpha=N$.
Depending on parameters $(\lambda_\alpha,\kappa_\alpha,
m^2)$ one obtains potentials of different forms. Since the potential
\eqref{potential-generic} is the sum of potentials which only depend
on a single field $\phi_\alpha$, we may analyze each of them
independently. The latter is basically the shape of the potential
about the single-block vacuum which is studied first. We then
analyze the multi-field case.

\subsection{Analysis of the potential around the single-block (irreducible) vacuum}
\label{single}

As a starter, we briefly review, summarize and expand on the results
of \cite{Ashoorioon:2009wa}, where Matrix Inflation theory is
effectively reduced to a single scalar field theory corresponding to
$\alpha=1$ in Eq. (\ref{general-solution}). For that purpose we assume
 the $N\times N$ matrix part of   $\Phi_i$ to be  proportional
to the generators of $SU(2)$ algebra in its $N\times N$
irreducible representation. Upon the rescaling
\be%
\hat\phi=\left[\frac14 N(N^2-1)\right]^{-1/2} \phi\ , %
\ee%
the action reduces to a canonically normalized scalar field theory
coupled to Einstein gravity with the potential%
\be
\label{VS}
V_{S}(\phi)=\frac{\lambda_{\phi}}{4} \phi^{4}-
\frac{2\kappa_{\phi}}{3} \phi^{3}+\frac{m^{2}}{2} \phi^{2} %
\ee%
where%
\be \lambda_{\phi} 
= \frac{8 \lambda}{ N (N^{2}-1)}  \ , \qquad \kappa_{\phi}
= \frac{2 \,\kappa}{\sqrt{N(N^{2}-1)}} \, . %
\ee%

\begin{figure}[t]
\includegraphics[angle=0, scale=0.55]{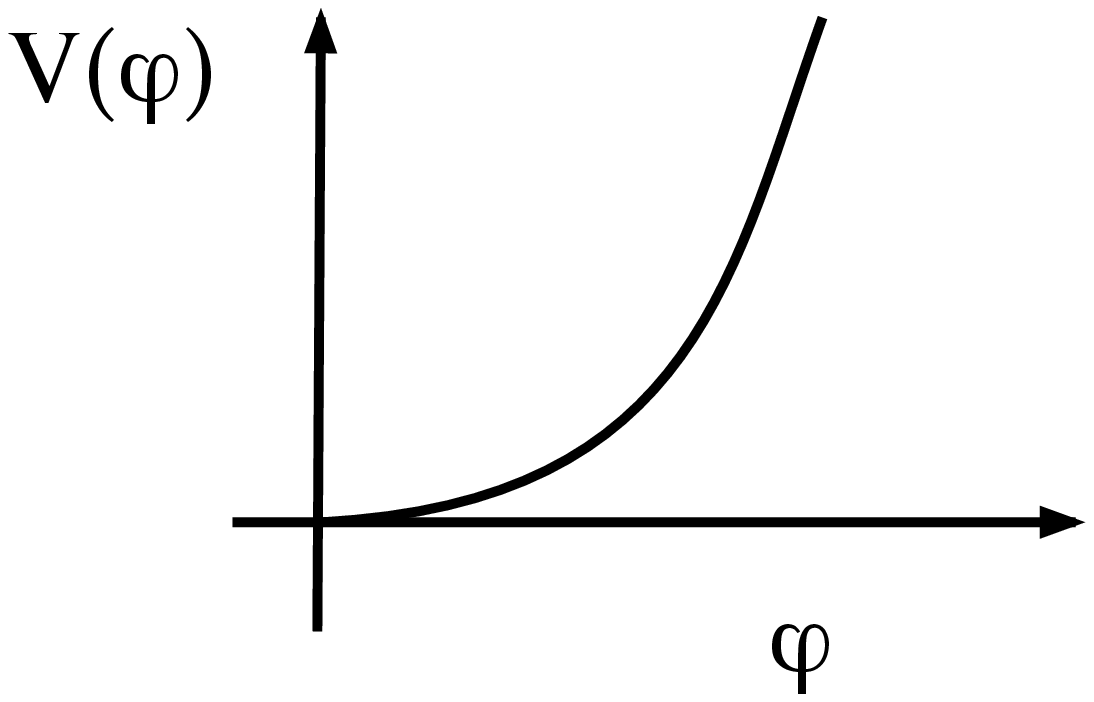}{\bf{(I)}}  \hspace{1cm}  \vspace{1cm}
\includegraphics[angle=0,scale=0.55]{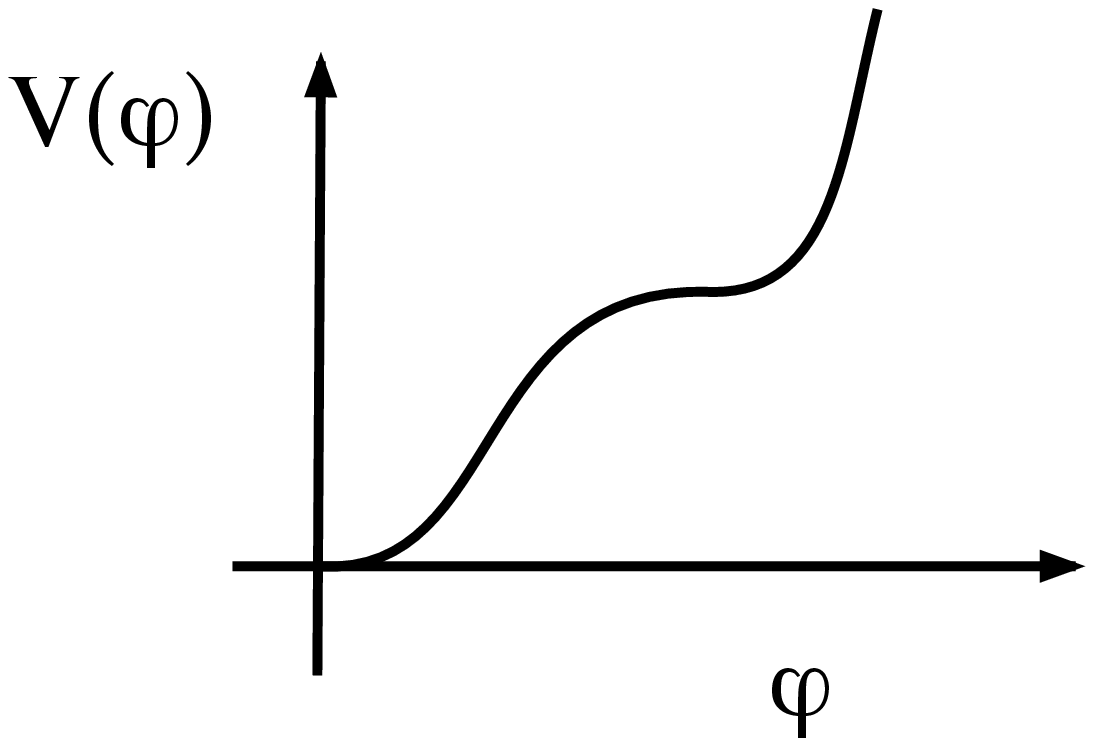}{\bf{(II)}} \vspace{1cm}
\includegraphics[angle=0, scale=0.55]{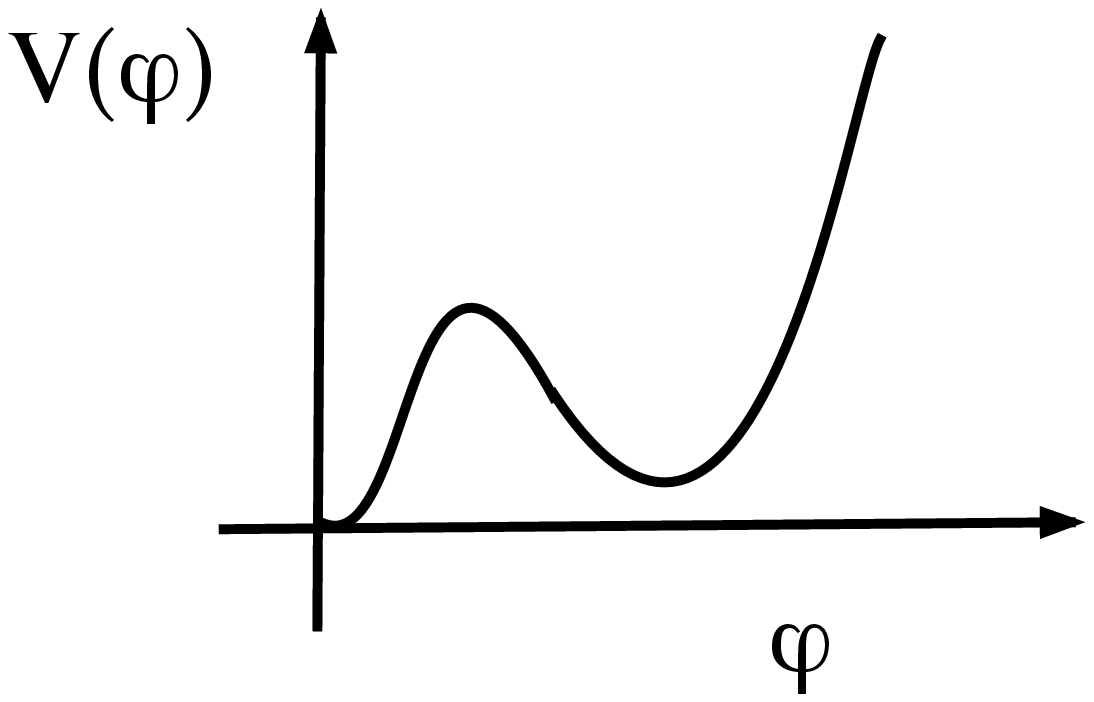} {\bf{(III)}} \hspace{1cm}
\includegraphics[angle=0,scale=0.55]{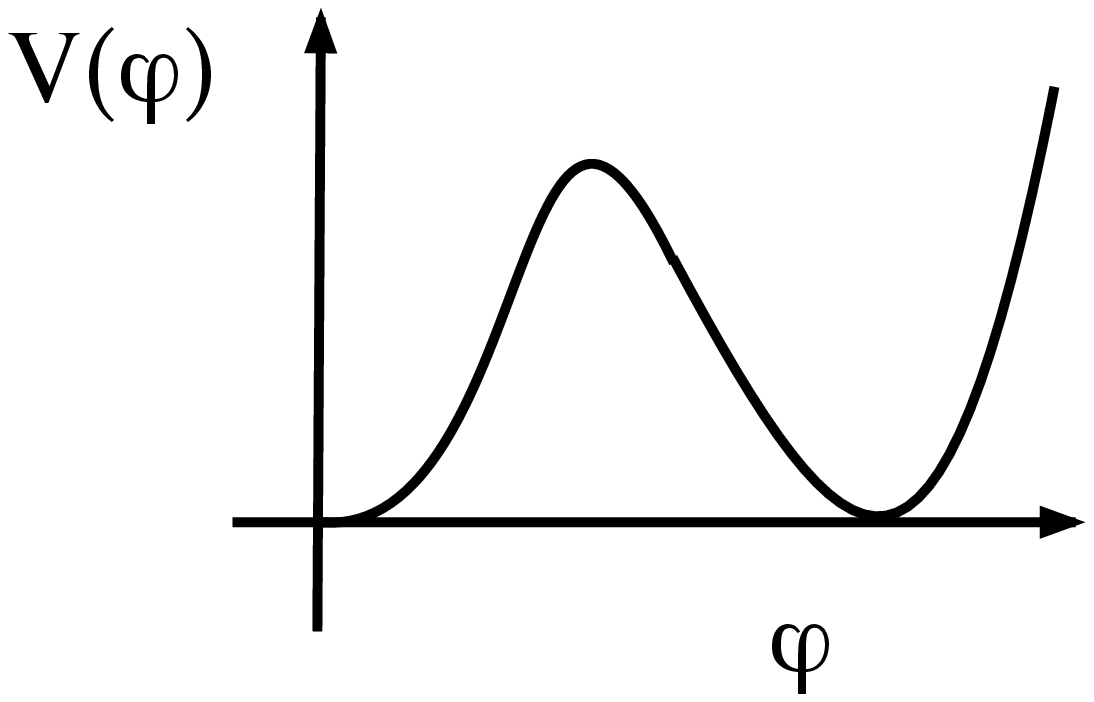}{\bf{(IV)}}
\includegraphics[angle=0,scale=0.55]{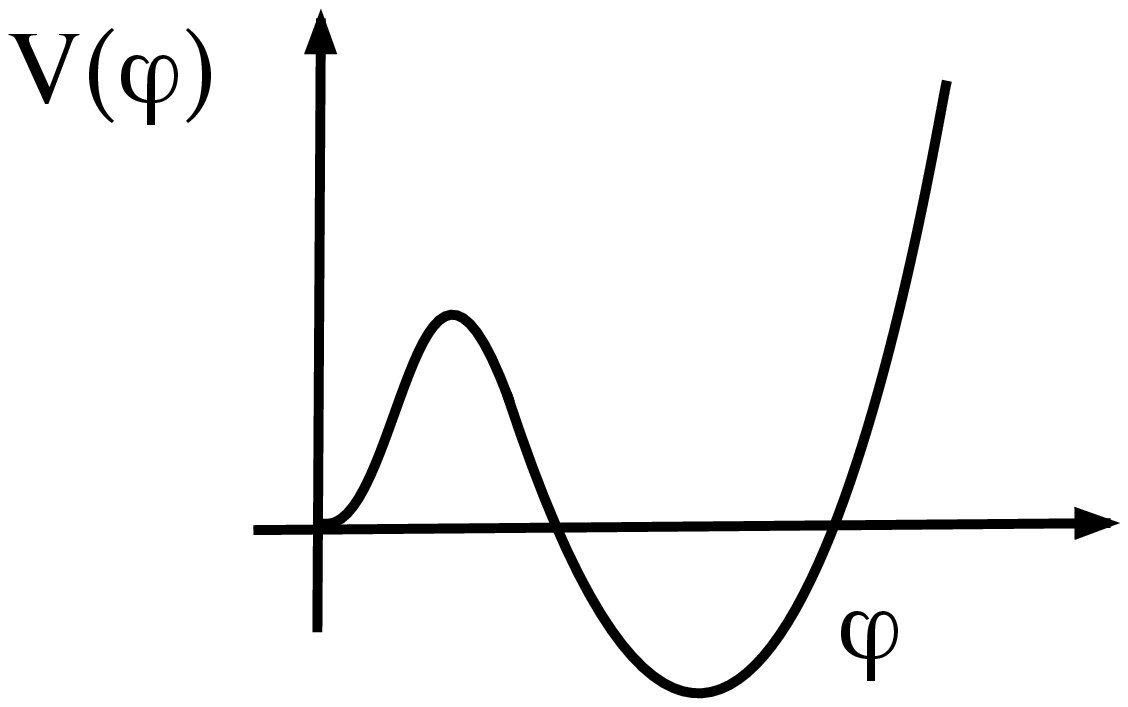}{\bf{(V)}}
\caption{Different possible shapes of the single block potential Eq.
(\ref{VS}). } \label{V-single}
\end{figure}

Depending on  the value of the ratio $m^2\lambda_\phi/\kappa^2_\phi=
2m^2\lambda/\kappa$ we may have five different cases.  Note the
remarkable property that this ratio is $N$-independent. In {\bf Fig. \ref{V-single}} we have
plotted the different shapes of the potential. Since in the string
theory setup, $\hat \phi_\alpha$ represents the radius of each fuzzy
sphere, we take $\phi_\alpha$ to be positive.

\begin{itemize}%

\item \textbf{Case I)} \quad $\kappa^2<2m^2\lambda$.

In this case the potential has only a single minimum at $\phi=0$ for
which the potential vanishes. All the $\kappa=0$ cases, including
$\lambda\phi^4$ and $m^2\phi^2$ chaotic inflation potentials are in
this class.

For the rest of cases $\kappa^2\geq 2m^2\lambda$ and hence one may use
the parameterization%
\be\label{Theta}%
\sin^2\Theta\equiv \frac{2m^2\lambda}{\kappa^2}\ \qquad \Theta\in [0,\pi/2]. %
\ee%

\item \textbf{Case II)}  \quad $\Theta = \frac{\pi}{2}$.

In this case the potential has an inflection point which happens at
$\phi_0=\frac{\kappa_\phi}{\lambda_\phi}$. The value of potential at
the inflection point is%
$$V_S^0=\frac{ \kappa^4}{384 \lambda^3}\ N(N^2-1)\ .$$%
As discussed in \cite{Ashoorioon:2009wa} the inflationary model
based on this case has a small red spectral index and is on the
verge of being ruled out.

\item \textbf{Case III)}  \quad     $  \sin^{-1}(\frac{8}{9})   <   \Theta <  \frac{\pi}{2} $.

In this case the potential besides the zero energy minimum at
$\phi=0$  has also a minimum at%
\be\label{minimum-phi}%
\phi_0=\frac{\kappa_\phi}{\lambda_\phi}(1+\cos\Theta)\ , %
\ee%
and the energy%
\be\label{minimum-energy}%
V_S^0=\frac{\kappa^4}{ 384 \lambda^3}\ N(N^2-1)\ F(\Theta)\ ,%
\ee%
where
$$ F(\Theta)=(1-3\cos\Theta)(1+\cos\Theta)^3.$$

In this case $V_S^0>0$. If the inflationary dynamics
happens around the $\phi_0$ minimum such that the end point of the
slow-roll inflationary phase is at $\phi_0$,  then the situation is like old inflation and we are back to the ``graceful exit'' problem. In  Appendix  B, we have determined the set of parameters for which nucleation rate via Coleman-De Luccia \cite{Coleman:1980aw} quantum tunneling is substantial to cause a first order phase transition at $\phi_0$. We have also shown that it is not possible to tunnel from the false vacuum to the true one by Hawking-Moss \cite{Hawking:1981fz} phase transition, if one demands the COBE normalization for density perturbations in the subsequent slow-roll phase.

\item \textbf{Case IV)}  \quad $  \Theta = \sin^{-1}(\frac{8}{9})  $.

In this case $\phi_0 = 4 \kappa_\phi/3 \lambda_\phi$ and
$F(\Theta)=0$ so the energy of the potential at the minimum
vanishes. In \cite{Ashoorioon:2009wa} this case was called  the
``symmetry breaking'' potential. This case does not suffer from the
secondary old inflation phase and hence is a preferred case for
building an inflationary scenario. Moreover, the potential in this
case has  the remarkable property that it could be obtained from a
cubic superpotential and hence the model can be embedded in a
supersymmetric theory. This latter gives us a better control over
the running of the parameters and the Coleman-Weinberg corrections.
In string theory set up, the minimum at $\phi_0$ corresponds to the
configuration when $N$ coincident D3-branes blow up into the
supersymmetric configuration of a giant fuzzy sphere with radius
determined by $\phi_0$. The other vacuum, $\phi=0$, corresponds to
the configuration of commuting matrices with a shrinking size
sphere.

\item \textbf{Case V)}  \quad $ 0\leq  \Theta < \sin^{-1}(\frac{8}{9})  $.

In this case $F(\Theta)<0$  and the global minimum is at $\phi_0$
with an AdS vacuum. The $m^2=0$ case, which is the case usually
studied in the D-brane setting in string theory (see e.g.
\cite{Myers:1999ps, Wadia, giant-inflaton}), falls in this class. It
is possible to have slow-roll inflation around the minimum $\phi=0$.
The minimum at $\phi_0$ is an AdS type and is not suitable as an end
point for inflation. As discussed in \cite{Coleman:1980aw} the gravitational effects makes the $\phi=0$ vacuum stable against tunneling to true vacuum at $\phi_0$.

\end{itemize}
In the specific examples of two-field models studied in the following section, we restrict our analysis mainly to the cases \textbf{I)}
and \textbf{IV)}  above.

\subsection{Analysis of the potential for $n$-block vacua}

We now consider a general $n$-block vacuum, corresponding to the
$n$-field inflationary potential \eqref{potential-generic} with
\eqref{lambda-kappa-scaling}.
In this case the equation of motion
for the fields $\phi_\alpha$ (when gravity is turned off) are
essentially decoupled from each other. Nonetheless, all of  them are
coupled to the background metric and give contributions to the Hubble
parameter and hence the field $\phi_\alpha$ will feel the effects of the
other fields through the evolution of the background metric. The
equations of motion for the
$n$-field theory are%
\bse \label{system-eq1}%
\begin{align}
 H^{2} &= \frac{1}{3 M_{P}^{2}} \sum_{\alpha=1}^n
\left( \frac{1}{2} \dot \phi_\alpha^{2} +
V_\alpha(\phi_\alpha) \right) \\
&\ddot \phi_\alpha + 3 H \dot \phi_\alpha + \partial_{{\phi_\alpha}}
V_\alpha=0\ ,
\end{align}%
\ese%
where $V_\alpha(\phi_\alpha)$ is  given by Eq. (\ref{potential-generic}).

Noting  that the angle $\Theta$ given by Eq. (\ref{Theta}) is
independent of $N_\alpha$, the form of the potentials $V_\alpha$ (in
the sense that which of the five cases of previous subsection they
fall into) is independent of $\alpha$ and hence for a generic
$n$-block vacuum we still have the same five cases discussed above.
The value of the potential at the minimum and at which $\phi_\alpha$
the minimum happens, however,  depend on $N_\alpha$.

In summary, irrespective of which set of $\{ N_\alpha \}$ is chosen, the
generic shape of the inflationary potentials falls into one of the
five cases discussed above and is completely specified by the
dimensionless ratio $\kappa^2/(m^2\lambda)$.

\section{Landscape of $n$-block Matrix Inflation}
\label{inflation-landscape}%

In the previous section we analyzed different possibilities for the
generic behavior of the inflationary potential depending on how the
original parameters of  M-flation is chosen. In this section we
study the landscape of various inflationary models arising from
inflation assuming a given  set of parameters $(\lambda,\kappa,
m^2)$ and $N$.

\subsection{Counting the number of vacua in the landscape}

As the first piece of information on the rich landscape of the
potentials, which all are in the generic form of
\eqref{potential-generic} with \eqref{lambda-kappa-scaling}, we
present a counting of these vacua. The case of single block, $n=1$, was studied
in \ref{single} and depending on the ratio
$m^2\lambda/\kappa^2$ we may have either of the five distinct cases
discussed earlier. Next, we may consider the double block vacuum,
which leads to a two-field inflationary potential. In this case, the
number of possibilities is the same as the ways one can partition $N$ into two
positive integers. Denoting the size of two blocks by $N_1$ and
$N_2$, $N_1+N_2=N$, with $N_1\geq \ N_2\geq 1$, one has $[\frac12 N ]$
possible solutions, where $[x]$ is the integer part of $x$. (Note
that exchanging $N_1$ and $N_2$ does not lead to a different
theory.) In other words, there are $[\frac12 N]$ possible two-field
models coming out of M-flation.

In a similar way one may count ${\cal N}_n(N)$, the number of
independent $n$-field inflationary models arising from M-flation
(which is the same as  the number of vacua or the minima of the
potential). ${\cal N}_n(N)$ is the number of ways an integer $N$ can
be partitioned into $n$ positive
integers $N_i$ such that $\sum_{i=1}^n N_i=N$, $1\leq N_i\leq N_{i+1}$. One can show that%
\be%
 {\cal N}_n(N)=\sum_{k=1}^{[\frac{N}{n}]} {\cal N}_{n-1}(N+n-1-n k) ,\qquad {\cal N}_1=1, %
\ee%
for $n\leq \frac{N}{2}$ and%
\be%
{\cal N}_{n}(N)={\cal N}_{N-n}(2(N-n))\ , %
\ee%
for $n\geq\frac{N}{2}$. From the above recursion relations one can compute ${\cal
N}_n(N)$. For example, for $n=2$, ${\cal N}_2=[\frac{N}{2}]$ and for
$n=N$, ${\cal N}_{N}=1$. For large $N$, and when $n\ll N$ one can
show that ${\cal N}_n(N)\sim \frac{1}{n!}(N-1)\cdots (N-n+1)\sim
\frac{1}{n!} N^{n-1}$.  The number of points (theories) in the
landscape of the
possible theories, for a given  set of parameters $(\lambda,\kappa,
m^2)$ and $N$ is %
\be\label{total}%
{\cal N}_{total}=\sum_{n=1}^N\ {\cal N}_n\sim \frac{2^N}{N}\ , %
\ee%
for large $N$.

It is worth noting that around each minimum point in this landscape we have an n-field
model specified with $\{N_\alpha\}$, where $\alpha=1,2,\cdots, n$. Around each point
there are $3N^2-n$ fields where $N=\sum_{\alpha=1}^{\alpha=n} N_{\alpha}$, that are not
classically excited. These fields, which are not classically coupled to the $n$
inflationary fields, can appear through quantum excitations of the fields, leading to
$3N^2-n$ isocurvature fields. In summary, among the cosmic perturbations around the
specific $n$-block vacuum, we have one usual curvature (adiabatic) mode, $n-1$ entropy
modes and $3N^2-n$ isocurvature modes.

\subsection{Transition between multi-field scenarios}
\label{quantum-landscape}

As  discussed, if we start with the initial condition that the
matrix valued fields $\Phi_i$ and their time derivative $\dot\Phi_i$
are given in terms of the  generic $SU(2)$ (reducible) representation, the
dynamics of M-flation is such that they always remain in the same
sector. In this sense the classical (inflationary) dynamics around
the ``multi giant vacua'' decouple from each other and one may build
an inflationary model around either of these.  If we start with a
field which is initially in the sector specified by a given set of
$N_\alpha$, then $N_\alpha$ remains a conserved quantity by the classical
trajectory of the system. In general various fields in the
same sector specified by a set of $N_\alpha$ can mix with each
other. That is, in general the inflationary trajectory in the
space of $\phi_\alpha$ is curved. We will discuss the details of the
inflationary dynamics for the special two-field case in the next
section.

This ``classical'' decoupling can, however, break due to quantum
effects and one can find various quantum paths connecting vacua with
different sets of $N_\alpha$. To see this let us suppose that the
ratio $m^2\lambda/\kappa^2$ takes a generic value and for the ease
of the analysis consider the epoch the inflationary dynamics has
ended and the fields are settled in the minimum of their potential.
Let us study the tunneling between  vacua corresponding to two
$SU(2)$ reducible representations specified by sets of
$N^{(1)}_\alpha$ and $N^{(2)}_\alpha$ which will respectively be
denoted by $J_i^{(1)}$ and $J_i^{(2)}$. One such path between the
two vacua is%
\be\label{transition-path} %
\Phi_i=\phi(t)J_i^{(1)}+(1-\phi(t))J_i^{(2)}\ ,\qquad \phi(t=0)=1,\
\phi(t=t_0)=0.%
\ee%
One may then evaluate and minimize the action for the above path and
use the WKB approximation to compute the transition amplitude by
exponentiating the value of the Euclidean action. (We comment that
there are other paths over which the action is smaller compared to
the one evaluated for the above path. In such cases these paths
would dominate the tunneling.)

The transition may happen from a local minimum with higher (or
equal) energy to a local minimum  with lower (or the same) energy.
To be specific let us compare the irreducible (single-giant) vacuum
with a generic $n$-giant vacuum. Depending on the value of the ratio
$m^2\lambda/\kappa^2$ the single or  $n$-giant vacua could have a
higher energy. It is
readily seen that%
\be%
V_{S}^0-V_{n-\rm{field}}^0=\frac{N^3-\sum_\alpha N_\alpha^3}{N(N^2-1)} V_S^0\  ,\qquad \sum_\alpha N_\alpha=N\ . %
\ee%
and hence $V_{S}^0-V_{n-\rm{field}}^0$ has the same sign as
$V_{S}^0$. Explicitly, for the case \textbf{III} the $n$-giant
vacuum is more stable and the single giant vacuum can decay (tunnel)
into the  multi-field vacua, as well as tunneling into its own
$\phi=0$ vacuum. For the case \textbf{V}, we have an opposite
situation and the single giant vacuum is the most stable one and the
multi-giant vacua can tunnel into the single giant one. Such an
analysis has been carried out in \cite{Wadia}. The symmetry breaking
case, case \textbf{IV}, is special in the sense that $V_S^0=0$ and
hence all the vacua have the same energy. The calculation of the
transition amplitude for this case has been carried out in some
detail in \cite{DSV}. In the analysis there it has been shown that
the path which minimizes the action is not of the form
\eqref{transition-path}. In \cite{DSV} it is shown that the height
of the potential between the two vacua grows like $N^2$ (for large
$N$), unlike the $N^3$ expected from a path like
\eqref{transition-path}.

\begin{figure}[t]
\includegraphics[angle=0, scale=.75]{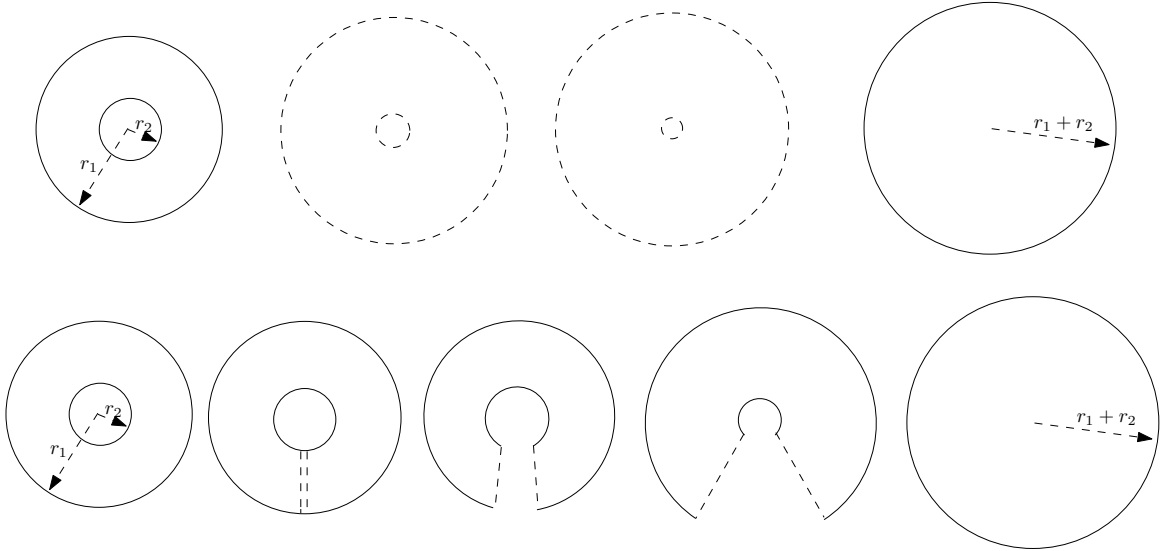}
\caption{Two possible paths for quantum tunneling between  two block
and  single block vacua. The top figure shows a paths in which the
spherical symmetry of the giants are preserved along the path. The
figure below shows a path in which the transition happens through
nucleation of a throat between the two giants, through which the
inner giant is eventually dissolved into the outer one. Both of
these paths are coming from  quantum fluctuations of the two initial
giants and can be understood as specific modes of open strings
stretched between the two giants. Analysis of \cite{DSV} shows that
the path depicted in the lower figure dominates the transition.}
\label{two-to-one}
\end{figure}

One may also provide a picture for the above tunneling in terms of
transition between a multi-giants configuration to a single giant
one. Let us consider the specific case of transition between two and
single giant configurations.  As was discussed the above reducible
solution with two blocks (in the large $N_1,N_2$ limit) has the
interpretation of concentric spherical D-branes whose radii,
$N_1\hat\phi_1/2$ and $N_2\hat\phi_2/2$  are changing in time.
Analysis of \cite{DSV} suggests a transition between the two is dominated by
the path which is mediated through nucleation of a throat between
the two giants, which eventually dissolves them into a single giant
(or vice versa), as depicted in \textbf{Fig. \ref{two-to-one}}.
These throats are basically the virtual open string loops which are
stretched between the giants (branes).

Moreover, there is also a non-zero tunneling amplitude from the
non-trivial ``giant'' graviton vacuum to the vacuum at $\phi=0$.  For the case of supersymmetric theories some of these paths may be forbidden due to supersymmetry.

\section{Two-Block Matrix Inflation}\label{two-block}

We have discussed that dynamics in a sector with a given set of
$N_\alpha,\ \alpha=1,2\ ,\cdots, n$ can decouple from the rest of
the theory if we initially start in that sector. Although the
inflationary potential in this sector contains $n$ decoupled
scalars, they are in general coupled to each other through
gravitational interactions, leading to  $n$-field inflationary
models.  In this section we analyze in some details the simplest of
such models, i.e. the two-field case for which the field $\Phi_i$
can be expanded as
\ba \label{initial}%
 \Phi_{i} = \hat \phi(t) J^{(1)}_{i} +
\hat \chi (t) J^{(2)}_{i} \, , %
\ea%
where $J^{(1)}_{i}$ and $J^{(2)}_{i}$ are respectively  $N_1$ and
$N_2$-dimensional ($N_1+N_2=N$) irreducible representations of
$SU(2)$ algebra. In the string theory setup studied in
\cite{Ashoorioon:2009wa}, the fields $\hat \phi$ and $\hat \chi$
indicate the radius of the giant spheres to which the stacks of
coincident branes are blown up. Since $\Phi_{i}$, $J^{(1)}_{i}$  and
$J^{(2)}_{i}$ are Hermitian, we conclude that $\hat \phi$  and $\hat
\chi$ are  real scalar fields.

After rescaling the $\hat\phi_\alpha$, $\lambda$ and $\kappa$ as in
\eqref{phi-scaling} and \eqref{lambda-kappa-scaling} we obtain the
effective two field  potential%
 \ba \label{Vphichi}
V_{D}(\phi, \chi)= \frac{\lambda_{\phi}}{4} \phi^{4} +
\frac{\lambda_{\chi}}{4} \chi^{4} - \frac{2\kappa_{\phi}}{3}
\phi^{3} - \frac{2\kappa_{\chi}}{3} \chi^{3}
   + \frac{m^{2}}{2} \left( \phi^{2} + \chi^{2} \right)\ .
\ea%
The effective couplings $\lambda_{\phi}, \lambda_{\chi},
\kappa_{\phi}$ and $ \kappa_{\chi}$ are given by
\ba \label{lameff}%
\lambda_{\phi} = \frac{8 \lambda}{ N_1 (N_1^{2}-1)} \ , \qquad
\kappa_{\phi} = \frac{2 \,
\kappa}{\sqrt{N_1(N_1^{2}-1)}} %
\ea%
with a  similar expressions for $\lambda_{\chi}, \kappa_{\chi}$ with
$N_1$ replaced by $N_2$.

With the initial condition of solutions given by Eq.
(\ref{initial}), our system is  basically reduced into a two-field
inflationary system of $\phi$ and $\chi$. The background equations
of motion are \ba \label{system-eq2} H^{2} = \frac{1}{3 M_{P}^{2}}
\left[  \frac{1}{2} \dot \phi^{2} + \frac{1}{2} \dot \chi^{2} +
V_D(\phi, \chi) \right] \nonumber\\
\ddot \phi + 3 H \dot \phi + \partial_{\phi} V_D=0 \quad , \quad
\ddot
\chi + 3 H \dot \chi + \partial_{\chi} V_D=0  \, .%
\ea%

Since $m^2\lambda_\phi/\kappa^2_\phi=m^2\lambda_\chi/\kappa^2_\chi$,
depending on the value of this ratio, the potential $V_D$ can be
classified into exactly the same five cases studied in subsection (\ref{single})
and hence we do not repeat them here. Below we study some
inflationary backgrounds constructed from $V_D(\phi, \chi)$. For some details of the treatment of the cosmological perturbation theory for
two-field models see  Appendix \ref{two-field}.

For general initial conditions, the inflationary trajectory is
curved in the $\phi$-$\chi$ plane. The general result of such an
effect will be a nonzero correlation between the adiabatic and the entropy perturbations. The magnitude of the effect,
however, depends on the values of the couplings and the initial
conditions. We consider the following cases, which shows to what
extent the results are initial condition-dependent.

\subsection{Quartic Potential}

Suppose $\kappa=m=0$. The background inflationary potential \ba V=
\frac{\lambda_{\phi}}{4} \phi^{4} + \frac{\lambda_{\chi}}{4}
\chi^{4} \, ,\ea%
has the form of case \textbf{I}.

Considering number of e-folds, $N_{e}$, as the clock $H dt = - d
N_{e}$, in the slow-roll limit one can check that
 \ba \phi \simeq \sqrt{8 N_{e}} M_{P} \sin
\alpha \quad , \quad  \ \chi \simeq \sqrt{8 N_{e}} M_{P} \cos \alpha
\, , \ea where $\alpha $ is an angle. Furthermore, the trajectory in
$(\phi, \chi)$ space or $(N_{e}, \alpha)$ space is \ba \label{traj-lp4}
 \phi^{-2} = R^{2} \chi^{-2} + C  \quad , \quad
 N_{e}= \frac{8 M_{P}^{2}}{C} \left(   \sin^{-2} \alpha  - { R^{2}} \cos^{-2} \alpha   \right)
\ea where $C$ is a constant of integration and $R$ is the ratio of
the couplings \ba R^{2} \equiv
\frac{\lambda_{\phi}}{\lambda_{\chi}} = \frac{N_2
(N_2^{2}-1)}{N_1(N_1^{2}-1)} \, . \ea The analysis is similar to
\cite{Polarski:1992dq, Polarski:1994rz} and \cite{Langlois:1999dw}.

For the numerical study, we first  consider the case where the
quartic coupling of $\chi$ and $\psi$ are of the same order. In this
case the trajectory is slightly curved, if the initial conditions of
none of the two fields are set to zero. For definiteness, we assume
that
 \ba
\lambda_{\phi} = 2.05\times 10^{-13} \quad, \quad   \lambda_{\chi} =5 \times 10^{-13} \nonumber\\
\phi_{i}=22.437~M_P \quad , \quad   \chi_i=8~M_P, \ea %
where the initial value of fields, $\phi_{i}$ and $\chi_i$, are
given $71$ e-folds before the end of inflation. Choosing the natural
bare value of $\lambda\sim 1$, from Eq. (\ref{lameff}), the above
values for the effective quartic couplings correspond  to the
following values for the dimensions of the blocks:
\begin{equation}\label{NM}
N_1\simeq 34000\quad, \quad N_2\simeq 25000 \, .
\end{equation}

\begin{figure}[t]
\includegraphics[angle=0,
scale=0.65]{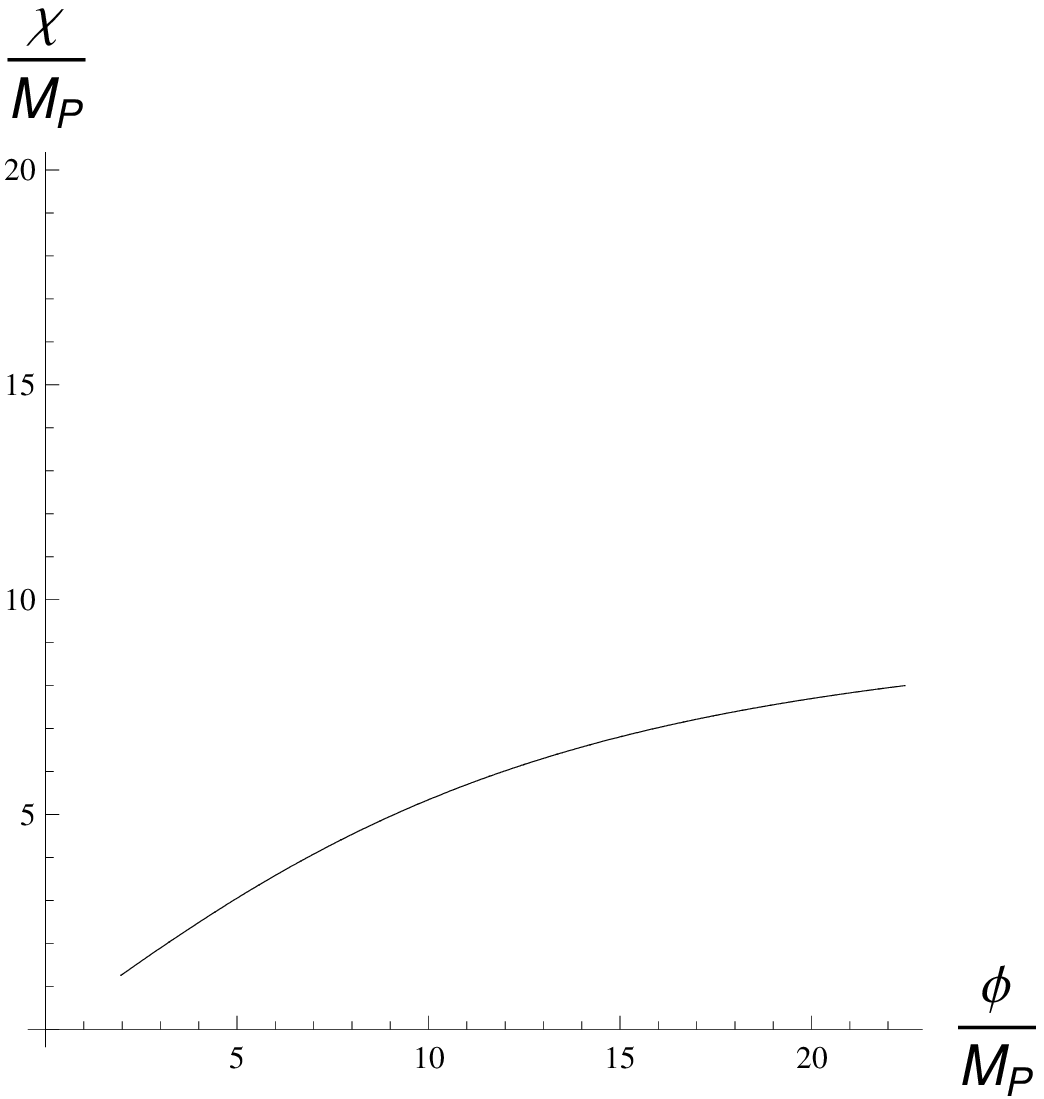}\hspace{2cm}
\includegraphics[angle=0,
scale=0.65]{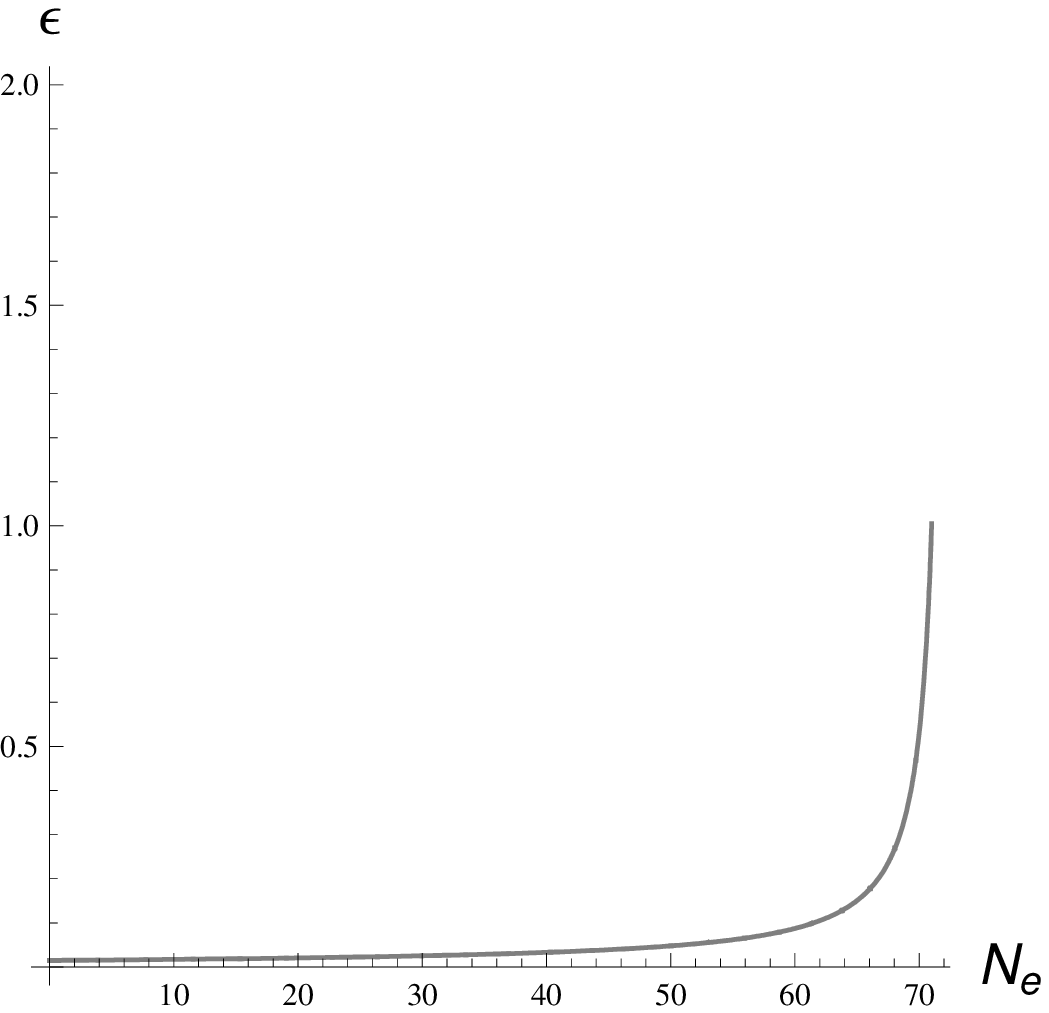} \caption{Left graph shows the
trajectory for two quartic potentials with comparable quartic
couplings. Right graph shows the slow-roll parameter, $\epsilon$, as
a function of number of e-folds. } \label{no-halt}
\end{figure}

The  trajectory in $\phi$-$\chi$ plane is given in left graph of
{{\bf Fig.\ref{no-halt}}}. It contains both the analytical curve
Eq. (\ref{traj-lp4}) and the exact numerical results which indicate
that the slow-roll approximation is very well valid until the end of
inflation. We have also graphed the first slow-roll parameter%
\ba
\epsilon= \frac{M_P^2}{2}\left(\frac{V_\sigma}{V}\right)^2\, , %
\ea%
in terms of number of e-folds in the right graph. As one can see,
$\epsilon$ changes smoothly up to the end of inflation. The value of
quartic couplings are chosen such that the amplitude of density
perturbations for the mode that exit $60$ e-folds before the end of
inflation, $k_{60}$, is COBE normalized, \textit{i.e.}
$P_{\mathcal{R}}(k_{60})=2.4\times 10^{-9}$. The scalar spectral
index at such scales is $n_{\mathcal{R}}(k_{60})\simeq 0.947$, which
is slightly smaller than single $\lambda \phi^4$ model, but still
within the $2\sigma$ level of WMAP results.

The amplitude of correlated entropy mode is $4.4\times 10^{-17}$ at
the current horizon scale and has a spectral index of $\simeq
0.9837$. The correlation between curvature and entropy
perturbations, $\mathcal{C}$, is $0.228$ at such scales which shows
that curvature perturbations at these scales are partially generated
through the transformation of entropy perturbations to curvature
ones. The amplitude of tensor perturbation is $P_T(k_{60})\simeq
6.33\times 10^{-10}$, corresponding to $r\simeq 0.26$ which is on the verge of being ruled out
by the future experiments. The tensor spectral index, $n_T$,
is $\simeq -0.036$. One should note that the consistency relation
between tensor and scalar spectra for single field inflation, $r=-8
n_T$, changes to $r=-8 n_T (1-\mathcal{C}^2)$ in such two-field
models \cite{Wands:2002bn}, which is confirmed by our numerics. In
this sense the two-field $\lambda\phi^4$ theory is subject to a
lower tensor-to-scalar ratio $r$ than the single field case. Hence
it may fall in the region of $n_s-r$ plane which is allowed by
WMAP5. (c.f. \cite{Ashoorioon:2005ep} for the violation of
consistency relation in the context of trans-Planckian physics).

\begin{figure}[t]
\includegraphics[angle=0,
width=70mm, height=75mm]{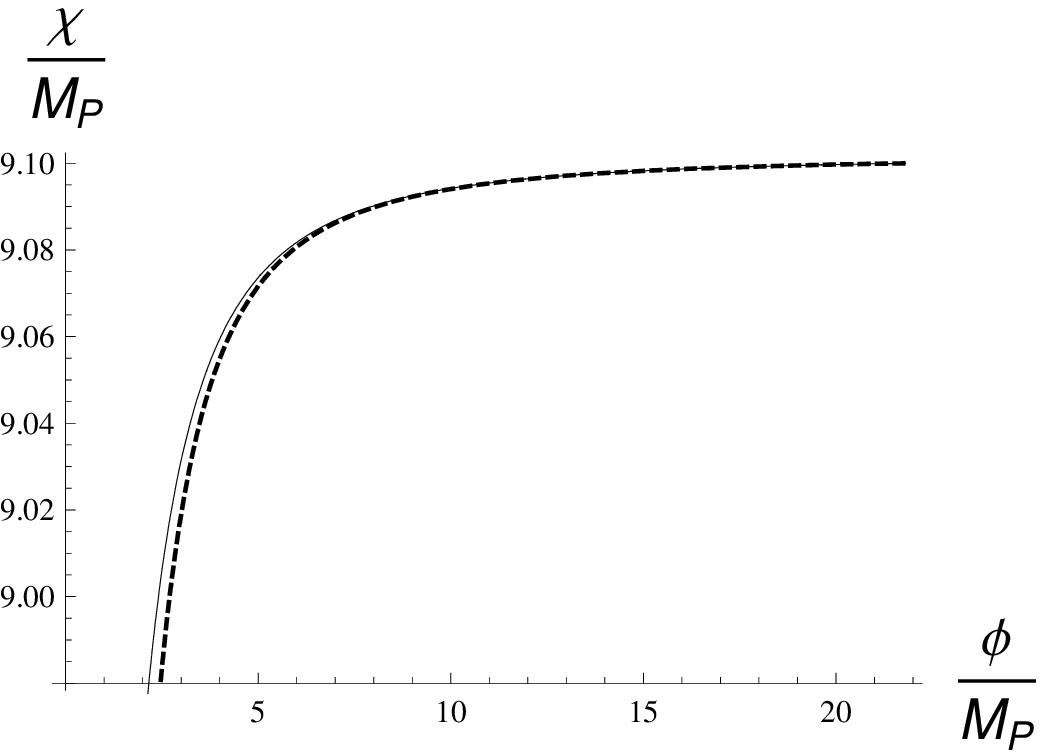} \hspace{2cm}
\includegraphics[angle=0,
scale=0.65]{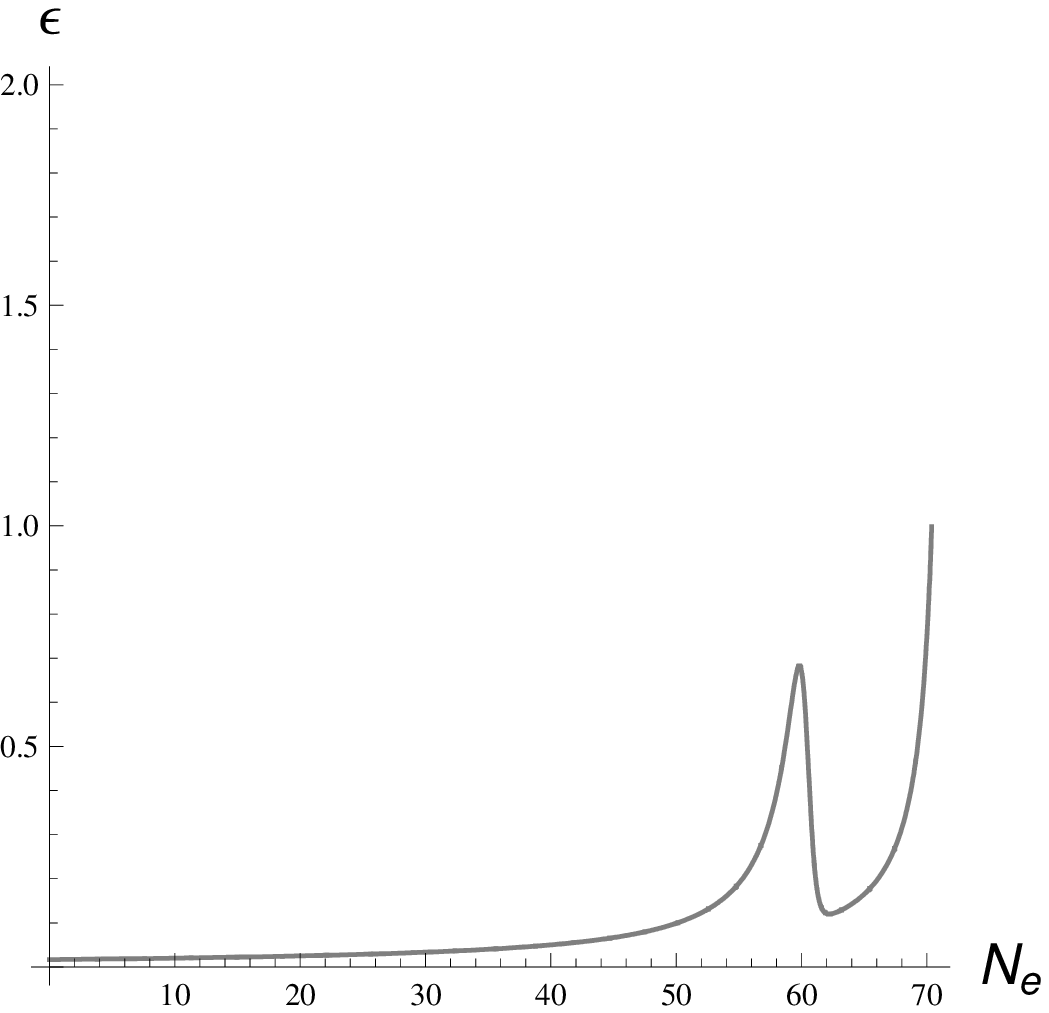} \caption{Left graph shows the
trajectory for two quartic potentials with hierarchy in quartic
couplings. There is a transient non-slow-roll phase in the evolution
of the fields. The solid and dashed curved, respectively, denote the
analytic and numerical results. Right graph shows the slow-roll
parameter, $\epsilon$, as a function of number of e-folds.}
\label{with-halt}
\end{figure}

In the other example that we will consider, there will be a
transient period of fast-roll evolution. To realize such a scenario
there should be a hierarchy between $\lambda_{\phi}$ and
$\lambda_{\chi}$. Inflation will not stop, nonetheless $\epsilon$
approaches close to unity for few e-folds. Before this transient
period, inflation is basically driven by one of the fields and after
that by the other. In $\lambda \phi^4$ model, the spectral index
depends on the number of e-folds, $N_e$, via the relation,
$n_{\mathcal{R}}=1-3/N_e$. Thus if the first phase lasts less than
$50$ e-folds, the scalar spectral index easily falls outside the
$2\sigma$ limit of WMAP5 central value for the spectral index.
Therefore double inflation models like \cite{Polarski:1992dq},
consistent with predictions in the CMB regime, are difficult to be
produced with two quartic potentials. Nonetheless, we investigate
this possibility that such a non-slow-roll phase occurs toward the
end of inflation. In the following example we have tuned the
parameters such that the non-slow-roll phase occurs in the last $10$
e-folds of inflation:
 \ba
\lambda_{\phi} = 2.05\times 10^{-13} \quad, \quad   \lambda_{\chi} = 5\times 10^{-16}\nonumber\\
\phi_{i}=21.8~M_P \quad , \quad   \chi_i=9.1~M_P\, .
 \ea
Initial values for the fields are given $70.34$ e-folds before the
end of inflation. The exact numerical trajectory is shown in the
left graph of {{\bf Fig.  \ref{with-halt}} by the solid curve,
whereas the analytic trajectory, given by formula (\ref{traj-lp4}),
is shown by the dashed one. The mismatch between the two curves is a
result of  slow-roll violation at the end of inflation. We have
also included the behavior of the  slow-roll parameter, $\epsilon$, vs.
$N_e$ which shows that for few e-folds $\epsilon$
increases considerably. Tuning the values of quartic
couplings such that the amplitude of density perturbations matches
the COBE normalization, one obtains  $n_{\mathcal{R}}(k_{60})\simeq
0.941$. The amplitude of correlated entropy perturbations are
$1.79\times 10^{-20}$ whose spectral index is $\simeq 0.958$. The
relative cross correlation between curvature and entropy
perturbations is $0.411$ which has a blue spectral index $\simeq
8.4\times 10^{-3}$. The amplitude of gravity waves at Hubble scale
is $\simeq 6.42 \times  10^{-10}$, \textit{i.e.} $r\simeq 0.328$,
which shows that this model is completely excluded by the
observations.


\begin{figure}[t]
\includegraphics[angle=0,
scale=0.65]{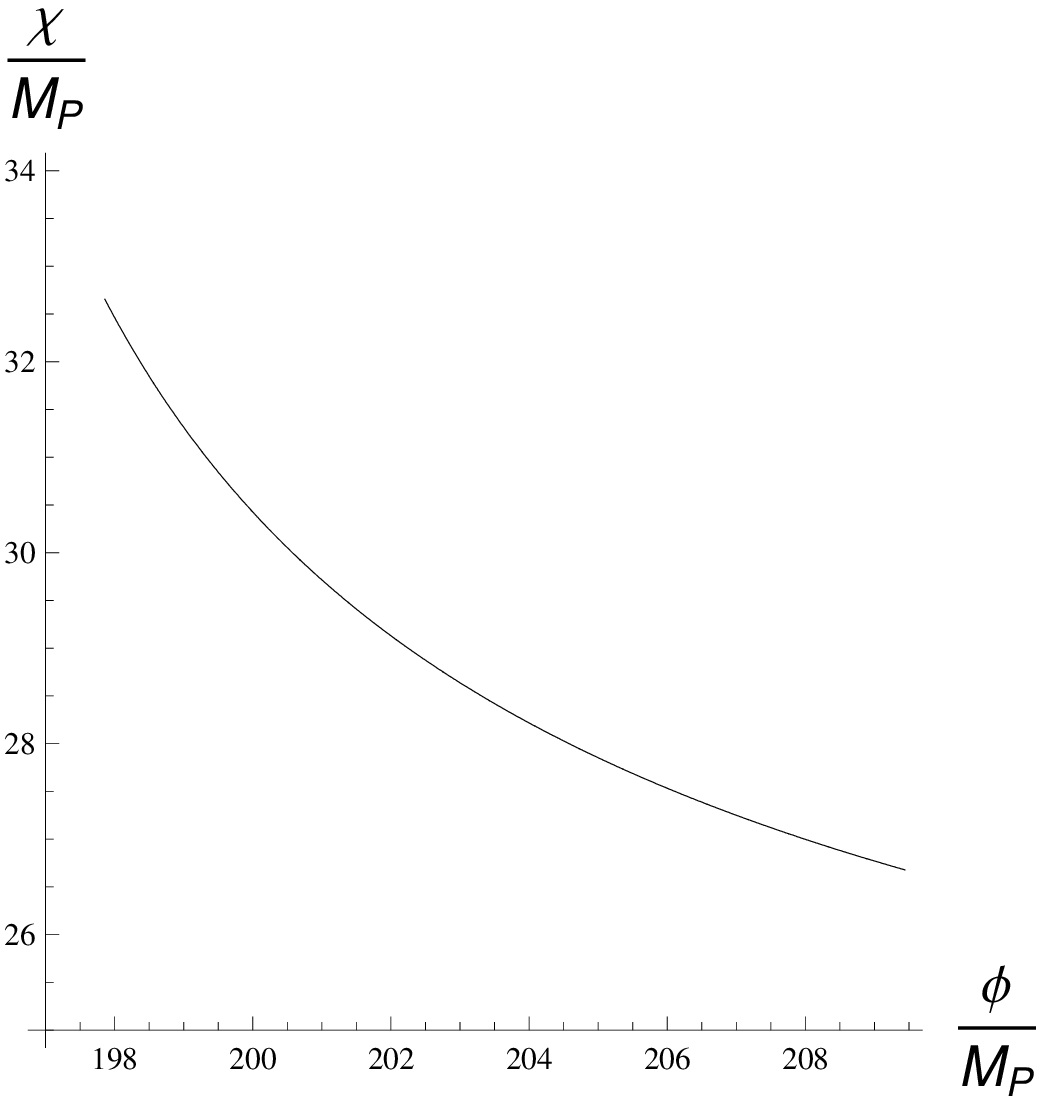} \hspace{2cm}
\includegraphics[angle=0,
scale=0.65]{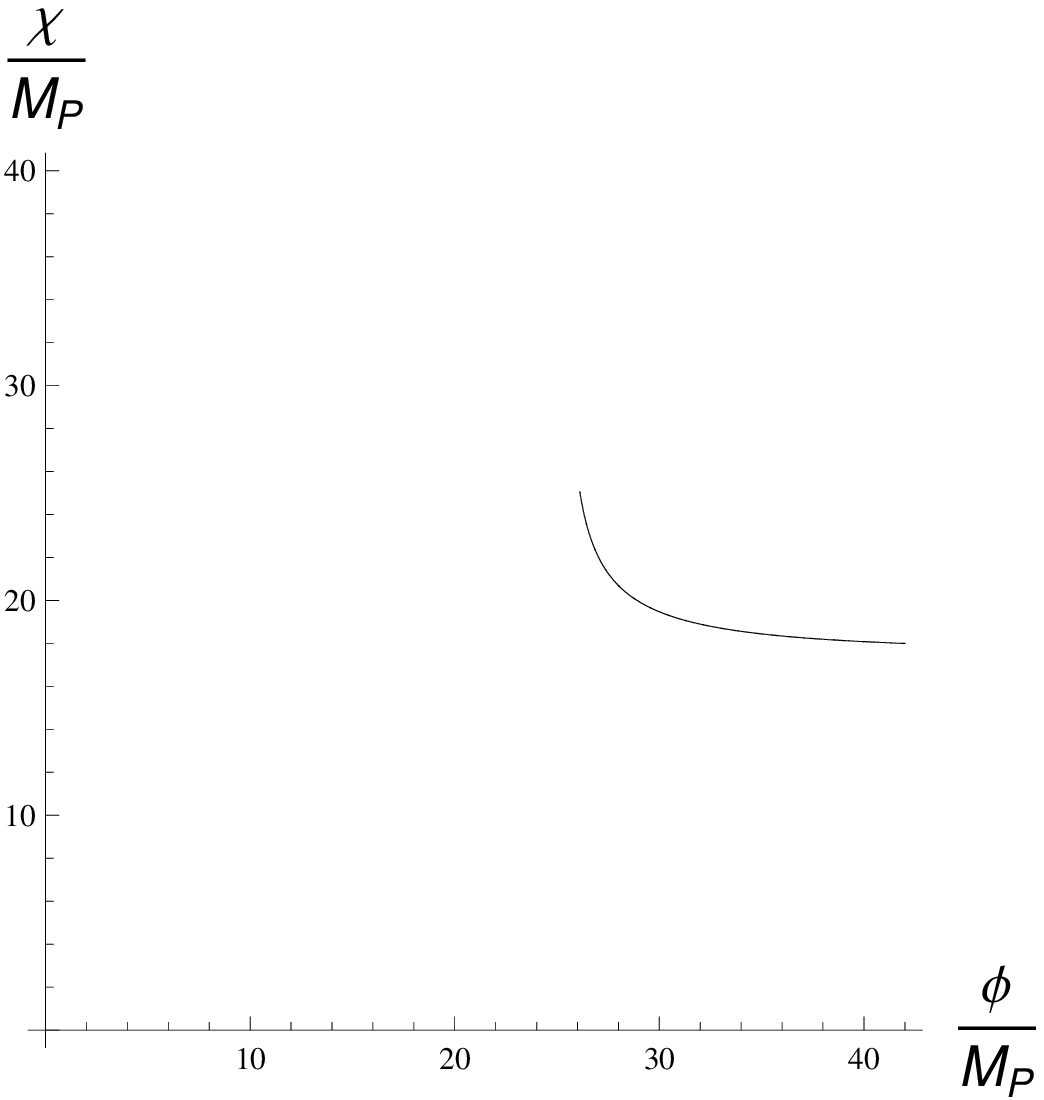}
\caption{ The graphs show the trajectory for two symmetry-breaking cases considered in the manuscript.}
\label{SB-traj}
\end{figure}

\subsection{Symmetry Breaking}
The next example we study corresponds to the symmetry breaking potential where
\ba
V=  \frac{\lambda_{\phi}}{4} \phi^{2}(\phi -
\mu_{\phi})^{2} +
 \frac{\lambda_{\chi}}{4} \chi^{2} (\phi - \mu_{\chi})^{2} \, .
\ea

Using the equations of motions for $\phi$ and $\chi$ combined
with Friedmann equation one can show that the curved inflationary trajectory in $\phi-\chi$
field space is
\ba
\label{sym-curve}
\phi(\phi -
\mu_{\phi}) + \chi(\chi -\mu_{\chi})  - \frac{\mu_{\phi}^{2}}{2}
\ln(2 \phi - \mu_{\phi}) -\frac{\mu_{\chi}^{2}}{2} \ln(2 \chi -
\mu_{\chi}) = 8 M_{P}^{2}  \, N_{e} \, . \ea

We consider the following values for the couplings:
\ba
 \lambda_{\phi} = 2\times 10^{-15}, \quad   \mu_{\phi} = 196.168 M_{P}, \quad \mu_{\chi}=36 M_{P}, \quad \lambda_{\chi}=\lambda_{\phi}{\left(\frac{\mu_{\phi}}{\mu_{\chi}}\right)}^2 \nonumber
\ea
\begin{equation}
    \phi_{i}=209.439~M_P \quad , \quad   \chi_i=26.678~M_P
\end{equation}
The trajectory in the $\phi-\chi$ plane is given in the left graph
of {\bf{Fig. \ref{SB-traj} }}. The theoretical curve Eq.
(\ref{sym-curve}) and the curve obtained from the   full numerical
analysis coincide with each other which indicates that the slow-roll
approximation is very well valid up to the end of inflation. For this
model, $n_{\mathcal{R}}\simeq 0.963$, which is compatible with WMAP
5 years results. The amplitude and spectral index of correlated
entropy mode are respectively $1.4\times 10^{-13}$ and $0.978$ at
horizon scales. The amplitude of correlation factor between
curvature and entropy mode is $\mathcal{C}\simeq 0.602$. For gravity
waves, the amplitude and spectral index are respectively,
$P_{T}(k_{60})=2.618\times 10^{-10}$, \textit{i.e.} $r\simeq 0.107$,
and $n_T\simeq -0.041$.

Another interesting case that has a nice geometric interpretation
is when $N_1=N_2$ and $\phi_i>\chi_i$. In the string theory picture, this corresponds to two stacks of D3-branes where in the background of RR six
form, $C_{(6)}$, two of their perpendicular dimensions are blown-up
to two concentric spheres, one of which has a radius smaller than
$\mu$ and the other one bigger than $\mu$. The bigger one shrinks
and the smaller one expands. These two spheres collide when their
radii reaches $\mu$ and pass through each other. This incident
happens well after the end of inflation and during preheating and
may have interesting consequences. One set of parameters for which,
the above scenario is realized is given below: \ba
 \lambda_{\phi} = \lambda_{\chi}=6\times 10^{-14}, \quad   \mu_{\phi} = \mu_{\chi}=26 M_{P},  \nonumber
\ea
\begin{equation}
    \phi_{i}=42~M_P \quad , \quad   \chi_i=18~M_P \, ,
\end{equation}%
where the initial conditions for the fields are given approximately
$73.28$ e-folds before the end of inflation. The trajectory in the
$\phi-\chi$ plane is given on the right graph of {\bf Fig.
\ref{SB-traj}}. Fixing the amplitude of curvature perturbations for
the modes that exit the horizon $60$ e-folds before the end of
inflation, one obtains the scalar spectral index,
$n_{\mathcal{R}}\simeq 0.954$. The amplitude and spectral index of
correlated entropy perturbations are respectively, $4.61\times
10^{-14}$ and $n_{\mathcal{S}}\simeq 0.963$. The relative cross
correlation between the corresponding entropy mode and curvature
perturbations is $\mathcal{C}\simeq 0.78$. The amplitude of tensor
perturbations is $3.16\times 10^{-10}$, \textit{i.e.} $r\simeq
0.13$.

\section{Discussions}\label{discussions}
In this work we continued the analysis of the Matrix Inflation
(M-flation) model we proposed in \cite{Ashoorioon:2009wa}. Our
analysis was focused along two directions. We first tried to map the
landscape of the inflationary models arising from M-flation. As
discussed, due to relative simplicity of our model compared to the
theories obtained from generic string theory (Calabi-Yau)
compactifications, we can give a complete map of the landscape, at
least classically. This enabled us to take first steps toward
studying quantum effects on the landscape. These are basically
tunneling (instanton effects in the scalar and/or gravity theories).
As we discussed, however, such transition amplitudes are too small
to bring us out of the local minimum. In this sense M-flation
provides a toy model with a fairly rich and at the same time
tractable landscape. This is more remarkable noting that M-flation
naturally and quite generically occurs in string theory setting as
low energy dynamics of D-branes in certain background fluxes.
However, in calculating the potentials from the dynamics of
coincident branes in an appropriate flux compactification, we
limited ourselves to low energy dynamics (lowest order in $\alpha'$)
and did not take into account the back-reactions of the
compactifications and the moduli stabilization on the potential
\cite{Kachru:2003sx, Burgess:2006cb, Baumann:2006th, Baumann:2007ah,
Chen:2008au}. These back-reactions can have significant effects.

 As we discussed if we start with a field configuration  for
which $\Phi_i$ are of the form \eqref{general-solution} our theory
(at classical level) effectively reduces to an $n$-field
inflationary model. If at $t=0$ we start with the sum of two such
solutions, which one leads to an $n_1$ field model and the other to
an $n_2$ field model with $n_1+n_2=N$, then the dynamics of the
theory is such that we will not simply get an  $n_1+n_2$ field model
and all $3N^2$ fields will eventually be turned on. It may, however,
happen that at different points in time one can approximate the
theory with an effective multi-field model, in which the number of
fields may change in time. This provides us with a situation similar
to the one discussed in \cite{Battefeld:2008qg}. It is interesting
to further study this line and examine the idea of ``meandering
inflation'' \cite{Tye:2009ff, Tye:2008ef} within our M-flation
setting which provides us with a tractable landscape.

Next, as a show case, we considered the two-field inflationary
scenarios which arise from M-flation and studied some of its specific
features, including the entropy modes, the spectral index $n_s$ and
the tensor-to-scalar ratio $r$.

Here we did not study the preheating scenarios which naturally arise
in a generic $n$-giant vacuum of M-flation. We expect the analysis
of the preheating and resonant particle creations by non-adiabatic
fluctuations of the scalar fields to be similar to the single-giant
case studied in \cite{Ashoorioon:2009wa}. The $n$-giant case,
however, is expected to have its own novel features too. For example,
in the corresponding spherical D-brane picture, one expects an
excessive particle production when giants (spherical branes) pass
through each other at the end of inflation \cite{Kofman:2004yc, McAllister:2004gd}.
When the branes become coincident some of the modes of open strings stretched between the
giants reach their minimum mass causing a resonant production of
these modes, leading to a very effective preheating scenario. This
point deserves further analysis which will be
carried out and presented elsewhere.

\vspace{1cm}


\textbf{\large{Acknowledgements }}

We would like to thank  Keshav Dasgupta, Brian Dolan, Liam McAllister,
Rob Myers, Herman Verlinde, Jiajun Xu and Henry Tye for valuable discussions
and comments. H. F. would like to thank the hospitality from Perimeter Institute and McGill University while this work was in progress. A.A. was supported by NSERC of Canada and MCTP, in the beginning of this project,
and the Uppsala University while it was being completed.

\appendix{}

\section{ Two Field Perturbation Theory}
\label{two-field}

Here we present the perturbation theory of two field models in some details.
In models with multiple inflaton fields, the field perturbations are decomposed into perturbation
tangential to the background inflationary trajectory, the adiabatic perturbation, and the perturbations orthogonal to the background trajectory, the entropy perturbations. For an extensive review see \cite{Wands:2007bd, Bassett:2005xm} and the references therein.

Following \cite{Gordon:2000hv}, the velocity in the field
space is $\dot{\sigma} \equiv \sqrt{\dot{\varphi}^2+\dot{\chi}^2}$
and we can define the polar angle in the field space as
\begin{equation}\label{theta}
 \cos\theta\equiv\dot{\varphi}/\dot{\sigma} \, .
\end{equation}
It is now useful to define the following Mukhanov-Sasaki variables:
\begin{equation}\label{Qsigma}
Q_{\sigma}=\cos\theta \, Q_{\varphi}+\sin \theta \, Q_{\chi} \; ,
\end{equation}
where
\begin{equation}\label{Qphi-Qchi}
Q_{\varphi}\equiv \delta\varphi+\frac{\dot{\varphi}}{H}\Phi~~~~~{\rm
and}~~~~~Q_{\chi}\equiv \delta\chi+\frac{\dot{\chi}}{H}\Phi \; .
\end{equation}
We work in the longitudinal gauge where in the absence of any anisotropic stress-energy tensor
the perturbed metric takes the following form \cite{Mukhanov:1990me}:
\begin{equation}\label{metric-long}
ds^2=-\big(1+2\Phi(\mathrm{t},\mathbf{x})\big)d\mathrm{t}^2 +
a(\mathrm{t})^2 \big(1-2\Phi(\mathrm{t},\mathbf{x})\big)
d\mathbf{x}^2.
\end{equation}

In the flat gauge, $Q_{\sigma}$ represents the field perturbations
along the velocity in the field space. $Q_{\sigma}$ is also related
to the commonly used curvature perturbation, $\mathcal{R}$, of the
comoving hypersurface via
\begin{equation}\label{R}
{\mathcal R}=\frac{H}{\dot{\sigma}}Q_{\sigma} \; .
\end{equation}
Similarly the isocurvature perturbation is:
\begin{equation}\label{s-Q}
\delta s= -\sin \theta \, Q_{\varphi}+\cos\theta \, Q_{\chi} \; .
\end{equation}
It describes field perturbation perpendicular to the field velocity
in the field space and, by analogy with $\mathcal{R}$, we can define
a rescaled entropy perturbation, $\mathcal{S}$, through
\begin{equation}\label{S}
\mathcal{S} = \frac{H}{\dot{\sigma}}\delta s \; .
\end{equation}
The transformations described above basically amount to introducing
a new orthonormal basis in the field space, defined by vectors
\begin{eqnarray}\label{vectors}
E_{\sigma}&=& (E_{\sigma}^\varphi,E_{\sigma}^\chi) = (\cos\theta,\sin \theta) \; ,\\
E_s&=&  (E_s^\varphi,E_s^\chi) = (-\sin\theta, \cos\theta) \; ,
\end{eqnarray}
which turn out to be useful to express various derivatives of the
potential with respect to the curvature and isocurvature
perturbations. Employing an implicit summation over the indices $I,J
\in \{\varphi,\chi\}$, one thus finds
\begin{equation}\label{VI}
V_{\sigma}=E^I_\sigma V_{I} \; , \qquad V_{s}=E_s^I V_I \; ,
\end{equation}
and
\begin{eqnarray}\label{VII}
V_{\sigma\sigma}=E^I_{\sigma}E^J_{\sigma}V_{IJ}\; , \qquad V_{\sigma
s}=E^I_{\sigma}E^J_{s}V_{IJ}\; , \qquad V_{ss}=E^I_{s}E^J_{s}V_{IJ}
\; .
\end{eqnarray}
for the first and second derivatives.

By combining the Klein-Gordon equations for the background scalar
fields one obtains the
background EOMs along the curvature and isocurvature directions
\begin{eqnarray}\label{BG-curv-iso}
\ddot \sigma+3H\dot{\sigma}+V_{\sigma}=0,\\
\label{dot-theta}
\dot{\theta}=-\frac{V_s}{\dot{\sigma}}.
\end{eqnarray}

With help of these equations, one can show that the EOMs for curvature and
isocurvature perturbations become
\begin{equation}\label{perturbations}
\ddot{Q}_{\sigma}+3H\dot{Q}_{\sigma}+\left(\frac{k^2}{a^2}+C_{\sigma\sigma}\right)Q_{\sigma}+
\frac{2V_s}{\dot{\sigma}}\dot{\delta s}+ C_{\sigma s}\delta s =0,
\end{equation}
\begin{equation} \label{perturbationsDelSig}
\ddot{\delta s}+3H\dot{\delta
s}+\left(\frac{k^2}{a^2}+C_{ss}\right)\delta
s-\frac{2V_s}{\dot{\sigma}}\dot{Q}_{\sigma}+C_{s\sigma}Q_{\sigma} =
0,
\end{equation}
with coefficients given by
\begin{eqnarray}
  C_{\sigma\sigma} &=& V_{\sigma\sigma}-{\left(\frac{V_s}{\dot{\sigma}}\right)}^2
  +\frac{2\dot{\sigma}V_{\sigma}}{M_{\rm P}^2 H}
  +\frac{3{\dot{\sigma}}^2}{M_{\rm P}^2}-\frac{{\dot{\sigma}}^4}{M_{\rm P}^4 H^2}\\
  C_{\sigma s} &=& 6 H \frac{V_s}{\dot{\sigma}}+\frac{2V_{\sigma} V_s}{{\dot{\sigma}}^2}+2V_{\sigma s}+
  \frac{\dot{\sigma} V_s}{M_{\rm P}^2 H} \\
  C_{ss} &=& V_{ss}- {\left(\frac{V_s}{\dot{\sigma}}\right)}^2\\
  C_{s\sigma} &=&-6H\frac{V_s}{\dot{\sigma}}-\frac{2
  V_{\sigma}V_s}{{\dot{\sigma}}^2}+\frac{\dot{\sigma} V_s}{M_{\rm P}^2 H} \; .
\end{eqnarray}

The power spectra of curvature (adiabatic), entropic perturbations and correlation spectrum are defined, respectively, as
\begin{equation}\label{power-spectra}
\mathcal{P}_{\mathcal{R}}(k)=\frac{k^3}{2\pi^2}\left\langle \mathcal{R}_{
\mathbf{k}}^{\star} \mathcal{R}_{\mathbf{k}'}\right\rangle
\delta^3(\mathbf{k}-\mathbf{k}') \; , \qquad \mathcal{P}_{\mathcal{S}}(k)=\frac{k^3}{2\pi^2}\left\langle \mathcal{S}_{\mathbf{k}}^{\star}
\mathcal{S}_{\mathbf{k}'}\right\rangle \delta^3(\mathbf{k}-\mathbf{k}')
\; .
\end{equation}
The correlation spectrum is defined as :
\begin{equation}\label{correlation-spectrum}
C_{\mathcal{R}\mathcal{S}}(k)= \frac{k^3}{2\pi^2} \left\langle \mathcal{R}_{
\mathbf{k}}^{\star} {\mathcal{S}}_{\mathbf{k}'}\right\rangle \delta^3(\mathbf{k}-\mathbf{k}')
\end{equation}
The correlation is also often quantified in terms of relative correlation coefficient, $\mathcal{C}$, which takes values between $0$ and $1$ and indicates to what extent final curvature perturbations result from interactions with entropy perturbations. It is defines as
\begin{equation}\label{rel-corr}
\mathcal{C}(k)=\frac{\left|\mathcal{C}_{\mathcal{R}\mathcal{S}}(k)\right|}{{\sqrt{\mathcal{P}_{\mathcal{R}}(k) \mathcal{P}_{\mathcal{S}}(k)}}}
\end{equation}

The curvature and isocurvature perturbations are evolved by
assuming initially, at conformal time $\tau_i$, a Bunch-Davies
vacuum. Therefore, when the wavelength of the two types of
perturbations is initially much smaller than the Hubble radius,
$k\gg aH$, we impose the initial conditions
\begin{equation}\label{initial conditions}
Q_{\sigma}(\tau_i)=\frac{e^{-ik\tau_i}}{a(\tau_i)\sqrt{2k}} \; ,
\qquad {\rm and} \qquad \delta
s(\tau_i)=\frac{e^{-ik\tau_i}}{a(\tau_i)\sqrt{2k}} \; .
\end{equation}
Inside the horizon these two modes are independent, because their
corresponding EOMs, eqs.~\eqref{perturbations} and
\eqref{perturbationsDelSig}, are independent in the limit $k \gg
aH$. However,  this does not hold
when the modes leave the horizon \cite{Tsujikawa:2002qx, Byrnes:2006fr}.

\section{Phenomenologically Viable Models in Case III}

In case {\bf III}, there is a nontrivial metastable giant vacuum at $\phi_0$ which has an energy larger than the true vacuum at $\phi=0$. If the primordial inflation occurs in the region $\phi>\phi_0$, it is possible that the inflaton gets trapped at the giant vacuum in $\phi_0$ and a secondary phase of old inflation occurs. This resurrects the graceful exit problem. However one may avoid this problem, if the universe can tunnel from the metastable vacuum. This could occur in two ways: through Hawking-Moss \cite{Hawking:1981fz} or Coleman-De Luccia \cite{Coleman:1980aw} tunneling. In the former, the universe can exit from the false vacuum expansion through a homogeneous bubble solution whose radius is greater than that of de Sitter space. The phase transition happens simultaneously everywhere and the universe jumps from its metastable vacuum to the maximum  of the potential and subsequently rolls downhill to its global minimum due to its tachyonic  perturbative mode. This occurs if $m^2\
 \leq 2 H^2$, where $m^2=\partial^2V/\partial\phi^2|\phi_0$. In the latter, which occurs when $m^2>2H^2$, there is a solution that interpolates between two vacua directly. Below we will show that transition from the false to the true vacuum is impossible, unless the height of bump is very small. In the Hawking-Moss case, the inflaton rolls downhill after the transition. Taking into account the COBE normalization for the amplitude of density perturbations for the modes that exit the horizon during the subsequent slow-roll phase in the region $\phi<\phi_0$, we show that Hawking-Moss phase transition is not possible for any value of parameters.

First let us calculate the nucleation rate with Coleman-De Luccia phase transition. The nucleation rate via the Colemand-De Luccia instanton, $\Gamma_{CD}$, is given as
\begin{equation}\label{Gamma}
\Gamma_{CD}=\mathcal{M}^4 \exp(-S_E),
\end{equation}
where $\mathcal{M}^2 \sim \mathcal{O}(V'')$ at the false vacuum and we approximate it with $V''$ hereafter. $S_E$ is the Euclidean action for the bounce solution that interpolates between the false and true vacua. Phase transition
can occur if \footnote{In this appendix we measure dimensionful parameters like $\kappa$ and $m$ in units of Planck mass and for the ease of notation set $M_P=1$.}
\begin{equation}\label{PT-cond}
\frac{\Gamma_{CD}}{H^4}\gtrsim\frac{9}{4\pi}
\end{equation}
To estimate the Euclidean action, we approximate the potential by a triangle and will use the results of \cite{Duncan:1992ai}. Following them, we denote the false and true vacua  as $\phi_+$ and $\phi_-$ with amplitudes $V_+$ and $V_-$, respectively. We also designate the local maximum  of the potential at $\phi_T$ with amplitude $V_T$. If
\begin{equation}\label{cond}
{\left(\frac{\Delta V_-}{\Delta V_+}\right)}^{1/2}\geq \frac{2\Delta\phi_-}{\Delta\phi_--\Delta\phi_+},
\end{equation}
we then have the following expression for Euclidean action:
\begin{equation}\label{Se1}
S_E=\frac{32\pi^2}{3}\frac{1+c}{{\left(\sqrt{1+c}-1\right)}^4}\left(\frac{{\Delta \phi_+}^4}{\Delta V_+}\right).
\end{equation}
In the above $\Delta\phi_{\pm}\equiv\pm (\phi_T-\phi)$, $\Delta V_{\pm}\equiv V_T-V_{\pm}$ and $c$ is the ratio of gradients of potential on either side of the local maximum:
\begin{equation}\label{}
c=\frac{\lambda_-}{\lambda_+},
\end{equation}
where
\begin{equation}\label{c}
\lambda_{\pm}=\frac{\Delta V_{\pm}}{\Delta \phi_{\pm}}.
\end{equation}
If condition \eqref{cond} is not satisfied, then the Euclidean action is given by the following expression:
\begin{equation}\label{Se2}
S_E=\frac{\pi^2}{96}\lambda_+^2 R_T^3 (-\beta_+^3+3c \beta_+^2 \beta_-+3c \beta_+\beta_-^2 -c^2 \beta_-^3),
\end{equation}
where,
\begin{equation}\label{betas}
\beta_{\pm}=\sqrt{\frac{8\Delta \phi_{\pm}}{\lambda_{\pm}}},
\end{equation}
and
\begin{equation}\label{Rt}
R_T=\frac{1}{2}\left(\frac{\beta_+^2+c\beta_-^2}{c\beta_--\beta_+}\right).
\end{equation}
Equipped with the following formulae, we calculate the Coleman-De Luccia nucleation rate in the case {\bf} above. Since,
\begin{eqnarray}
  \phi_T &=& \frac{\kappa_{\phi}}{\lambda_{\phi}}(1-\cos\Theta) \\
   \phi_+&=&  \frac{\kappa_{\phi}}{\lambda_{\phi}}(1+\cos\Theta) \\
   \phi_-&=& 0
\end{eqnarray}
Inequality \eqref{cond} is satisfied only when $\Theta$ does not fall in the interval $\left.\left(\arccos(\frac{1}{3}), 1.401496956\bar{5}\right.\right]$. Inside this interval, the Euclidean action is given by expression \eqref{Se2} which yields:
\begin{equation}\label{Se2-results}
S_E=\frac{f_1(\Theta)}{\lambda_{\phi}},
\end{equation}
where $f_1(\Theta)$ is a complicated function, but it is enough to know that it is monotonically decreasing as a function of $\Theta$ and starts from infinity at $\arccos(\frac{1}{3})$ and decreases to $344.47\bar{7}$ at $1.401496956\bar{5}$. Since $\lambda_{\phi}\leq 1$, the nucleation rate remains negligible in this interval of $\Theta$.

Outside this interval, the Euclidean action is given by expression \eqref{Se1}:
\begin{equation}\label{Se1-theta}
S_E=\frac{f_2(\Theta)}{\lambda_{\phi}},
\end{equation}
where
\begin{equation}\label{f}
f_2(\Theta)=\frac{128 \pi^2 (1+\frac{3 \cos^3\Theta-5 \cos^2\Theta+\cos\Theta+1}{8\cos^2\Theta})}{{\left(\sqrt{1+\frac{3 \cos^3\Theta-5 \cos^2\Theta+\cos\Theta+1}{8\cos^2\Theta})}-1\right)}^4}.
\end{equation}
$f_2(\Theta)$ reaches zero when $\Theta$ tends to $\frac{\pi}{2}$. Close to $\frac{\pi}{2}$, $f_2(\Theta)$ behaves like
\begin{equation}\label{f2}
f_2(\Theta)\sim -1024 \pi^2\left[{\left(\Theta-\frac{\pi}{2}\right)}^3+\cdots\right]
\end{equation}
Mass parameter and Hubble parameter at $\phi_+$ can be calculated and they turn out to be as follows:
\begin{equation}\label{mplus}
\mathcal{M}^2\simeq m(\phi_+)^2=\frac{2\kappa^2 \cos\Theta (1+\cos\Theta)}{\lambda_{\phi}},
\end{equation}
\begin{equation}\label{Hplus}
H^2(\phi_+)=\frac{\kappa^4 (1+\cos\Theta)^3 (1-3\cos\Theta)}{36\lambda_\phi^3}.
\end{equation}
Plugging these expressions into eq. \eqref{PT-cond} and expanding its L.H.S. around $\Theta=\frac{\pi}{2}$, one obtains
\begin{equation}\label{lhs-expanded}
\frac{\Gamma_{CD}}{H^4}\approx\frac{5184 \lambda_{\phi}^4(\Theta-\frac{\pi}{2})^2}{\kappa_{\phi}^4}\exp\left(\frac{1024\pi^2(\Theta-\frac{\pi}{2})^3}{\lambda_{\phi}}\right).
\end{equation}
If we call $x\equiv-\frac{1024\pi^2(\Theta-\pi/2)^3}{\lambda_{\phi}}$, eq.\eqref{PT-cond} takes the form
\begin{equation}\label{GoverH}
\frac{81 \lambda_{\phi}^{14/3}}{{(2\pi)}^{2/3}\kappa_{\phi}^4}x^{2/3}\exp(-x)\gtrsim\frac{9}{4\pi}.
\end{equation}
Noting that $0\leq\Theta\leq \pi/2$, $x$ varies between $0$ and $\frac{128 \pi^5}{\lambda_{\phi}}$. This equation has solutions, if the maximum of L.H.S. of equation is bigger or equal to $9/4\pi$. Since the maximum of L.H.S. of equation happens at $x=2/3$, we obtain the following constraints on the parameters:
\begin{equation}\label{const}
\frac{\lambda_{\phi}^{14/3}}{\kappa_{\phi}^4}\geq \frac{\pi^{1/3}3^{2/3}}{36}\exp(2/3)\simeq 0.1648.
\end{equation}
If this condition is satisfied, since the L.H.S. of the equation has a Maxwellian shape, in general we will have two solutions $x_1$ and $x_2$ for the equation which correspond to angles $\Theta_1$ and $\Theta_2$ where $\frac{\pi}{2}-\Theta_i=\left(\frac{\lambda x_i}{1024 \pi^2}\right)^{1/3}$. For
\begin{equation}\label{Valid-Theta}
\left(\frac{\lambda x_1}{1024 \pi^2}\right)^{1/3}\leq \frac{\pi}{2}-\Theta \leq \left(\frac{\lambda x_2}{1024 \pi^2}\right)^{1/3},
\end{equation}
the Coleman-De Luccia nucleation rate is bigger than the critical rate and inflation can end via Coleman-De Luccia first order phase transition. On the other hand, Coleman-De Luccia solution only exist when $m^2>2H^2$. This  results in addition constraint on the couplings
\begin{equation}\label{CDL-theta}
\frac{\kappa_{\phi}^2}{\lambda_{\phi}^2}<\frac{36\cos\Theta}{(1+\cos\Theta)^2(1-3\cos\Theta)},
\end{equation}
or since $\Theta$ is very close to $\pi/2$
\begin{equation}\label{CDL-theta-pi2}
\frac{\pi}{2}-\Theta<\frac{\kappa_\phi^2}{36\lambda_\phi^2}
\end{equation}
If the parameters $\kappa_{\phi}$ and $\lambda_{\phi}$ are tuned such that $\Theta$  satisfies \eqref{CDL-theta-pi2} and \eqref{Valid-Theta} simultaneously, we will have successful Coleman De-Luccia phase transition at the end of slow-roll inflation. The potential expanded around $\phi=\phi_+$, with $\Theta\simeq \frac{\pi}{2}$, is:
\begin{equation}\label{pot-around-pi_+}
V(\varphi)=\frac{\lambda_{\phi}}{4} \varphi^4+\frac{\kappa_{\phi}}{3}\varphi^3+\frac{\kappa_{\phi}^4}{12\lambda_{\phi}^3},
\end{equation}
where $\varphi=\phi-\phi_+$. For $\varphi$'s bigger than but close to zero $\eta$, the second slow-roll parameter, is bigger than one. However for $\varphi\sim {\cal O}(10) M_{P}$, the potential can sustain chaotic inflation. It is possible to show that one can satisfy the above constraints and the ones from WMAP simultaneously. For example consider the following values for $\lambda_{\phi}$ and $\kappa_\phi$
\begin{eqnarray*}
  \lambda_{\phi} = 1.7\times 10^{-18} , \qquad
  \kappa_{\phi} =2.5\times 10^{-21}~M_P
\end{eqnarray*}
The constraint \eqref{const} is clearly satisfied and one finds the following allowed interval from combining \eqref{Valid-Theta} and \eqref{CDL-theta-pi2}:
\begin{equation}\label{valid-theta-example}
2.694\times 10^{-8}\leq \frac{\pi}{2}-\Theta<6.007\times 10^{-8}.
\end{equation}
For these values of couplings the amplitude of density perturbations $60$ e-folds before the end of inflation matches the COBE normalization, $2\times 10^{-5}$ and the spectral index is $0.9508$ which is within the WMAP5 $2\sigma$ level.

We also would like to show that the phase transition from the false vacuum to the true one cannot happen via Hawking-Moss phase transition, if one demands to produce the observed amplitude of density perturbations in the subsequent slow-roll phase. The nucleation rate for Hawking-Moss phase transition is given by an expression similar to \eqref{Gamma} where now the Euclidean action is calculated for the Euclidean solution that interpolates between $\phi_+$ and $\phi_T$. According to \cite{Hawking:1981fz}, the Euclidean action is given by
\begin{equation}\label{S-Hawking-Moss}
S_{HM}=8\pi^2\left(\frac{1}{V_+}-\frac{1}{V_T}\right)
\end{equation}
or
\begin{equation}\label{SHM-Theta}
S_{HM}=\frac{1536 \pi^2 \lambda_{\phi}^3\cos^3\Theta}{\kappa_\phi^4\sin^6\Theta (1-3\cos \Theta)(1+3\cos\Theta)}.
\end{equation}
$S_{HM}$ approaches zero as we approach $\Theta=\frac{\pi}{2}$. This justifies that we expand it around this point:
\begin{equation}\label{SHM-expanded}
S_{HM}=-\frac{1536 \pi^2\lambda_\phi^3}{\kappa_\phi^4}\left(\Theta-\frac{\pi}{2}\right)^3+\cdots
\end{equation}
Using \eqref{mplus} \& \eqref{Hplus}, we have
\begin{equation}\label{GHM-H4}
\Gamma_{HM}/H^4\approx\frac{5184\lambda_\phi^4(\Theta-\frac{\pi}{2})^2}{\kappa_\phi^4}\exp\left[\frac{1536\pi^2\lambda_{\phi}^3}{\kappa_\phi^4}\left(\Theta-\frac{\pi}{2}\right)^3\right]
\end{equation}
We now follow the same procedure we did to analyze the behavior of the nucleation rate in the case of Coleman-De Lucica phase transition. We define the variable $x$ in the following manner
\begin{equation}\label{x-theta}
x\equiv-\frac{1536\pi^2\lambda_\phi^3}{\kappa_\phi^4}\left(\Theta-\frac{\pi}{2}\right)^3,
\end{equation}
In terms of which the equation $\Gamma_{HM}/H^4\gtrsim 9/4\pi$ takes the form
\begin{equation}\label{}
\frac{3^{10/3}\lambda_\phi^2}{\kappa_{\phi}^{4/3}\pi^{2/3}}x^{2/3}\exp(-x)\gtrsim\frac{9}{4\pi}.
\end{equation}
In order to have a non-empty set of solution for this equation, the couplings have to satisfy the condition
\begin{equation}\label{lambda-kappa-HM}
\frac{\lambda_\phi^2}{\kappa_\phi^{4/3}}\geq \frac{9}{4\pi}\exp(2/3)\left(\frac{\pi^2}{26244}\right)^{1/3}\simeq 0.1.
\end{equation}
This condition on the parameters is necessary to have successful phase Hawking-Moss phase transition from $\phi_+$ to $\phi_T$. If one requires that the inflationary scenario that arises in the region $\phi \leq \phi_T$ be phenomenologically viable too, one should also take into account the constraint that amplitude of density perturbations sets on the parameters \eqref{const}. The amplitude of density perturbations $60$ e-folds before the end of inflation is set by the ratio $\frac{\lambda_{\phi}}{m}$:
\begin{equation}\label{density}
\delta_H\simeq \frac{2}{5\pi}\frac{\lambda_{\phi}}{m} N_e^2\simeq 2\times 10^{-5}
\end{equation}
where $N_e$ is set equal to $60$ which corresponds to the moment CMB scales have left the horizon. This equation sets the following relation between $m$ and $\lambda$
\begin{equation}\label{lambda-m}
m\simeq 2.29\times 10^7 \lambda_{\phi},
\end{equation}
Since $\Theta$ is very close to $\frac{\pi}{2}$, the condition necessary to have inflection point inflation, $\kappa_{\phi}=m\sqrt{\lambda_{\phi}}$, still holds to a very good approximation, i.e.
\begin{equation}\label{kappa-lambda}
\kappa_\phi\simeq 2.29\times 10^7 \lambda_{\phi}^{3/2}.
\end{equation}
 Plugging this condition into inequality \eqref{lambda-kappa-HM}, one obtains the L.H.S. to be just a number, i.e. $1.53\times 10^{-10}$, which is much smaller than the R.H.S. of inequality. This means that Hawking-Moss phase transition with a subsequent inflationary period with the correct amplitude of density perturbations is not possible. This might be problematic for MSSM inflation \cite{Allahverdi:2006iq}, since it shows that in the case one deviates slightly from inflection point condition such that the potential acquires a metastable minimum at $\phi_0\neq 0$, the probability of Hawking-Moss tunneling from the false vacuum to the maximum of the potential is small, if one requires to obtain the correct amplitude for density perturbations in the subsequent slow-roll phase. This is contrary to the result of \cite{Allahverdi:2006we}, which had claimed otherwise. In fact, \cite{Allahverdi:2006we} had only proven the existence of Euclidean instantonic solution that extrapola
 tes between the metastable minimum and maximum of the potential and had not calculated the rate by which this tunneling occurs. Our computation, comparing the nucleation and the expansion rates, excludes such a possibility.


\end{document}